\pdfoutput=1

\PassOptionsToPackage{table,dvipsnames}{xcolor}
\documentclass[11pt]{article}
\usepackage{enumitem}
\usepackage{framed} 
\usepackage[final]{acl}
\usepackage{longtable} 
\usepackage{tabularx} 
\usepackage{tcolorbox}
\usepackage{float}
\usepackage{placeins} 
\renewcommand{\arraystretch}{1.3} 
\usepackage{times}
\usepackage{latexsym}
\usepackage{booktabs}   
\usepackage{tabularx}    
\usepackage{graphicx}

\usepackage[T1]{fontenc}

\usepackage[utf8]{inputenc}

\usepackage{microtype}

\usepackage{inconsolata}
\usepackage{listings}
\usepackage{amsmath}
\usepackage{graphicx}
\usepackage{booktabs}
\usepackage{multirow}
\usepackage{makecell}
\newcommand{\sysname}{\textit{Insight}Agent}

\definecolor{mygreen}{RGB}{0,150,0}
\newcommand{\preferred}[1]{{\color{mygreen}#1}}
\newcommand{\lesspreferred}[1]{{\color{orange}#1}}
\newcommand{\notpreferred}[1]{{\color{red}#1}}
\title{
Completing A Systematic Review in Hours instead of Months with Interactive AI Agents
}

\author{
  Rui Qiu\textsuperscript{1 *}\quad 
  Shijie Chen\textsuperscript{1 *}\quad 
  Yu Su\textsuperscript{1}\quad 
  Po‑Yin Yen\textsuperscript{2}\quad 
  Han‑Wei Shen\textsuperscript{1} \\[-0.5ex]
  \textsuperscript{1}The Ohio State University \quad
  \textsuperscript{2}Washington University School of Medicine \\[-0.5ex]
  {\texttt{
    \{qiu.580, chen.10216, su.809, shen.94\}@osu.edu, yenp@wustl.edu}
  }
}

\begin{document}
\maketitle
\def\thefootnote{*}\footnotetext{Equal contribution.}\def\thefootnote{\arabic{footnote}}
\begin{abstract}

Systematic reviews (SRs) are vital for evidence-based practice in high stakes disciplines, such as healthcare, but are often impeded by labor-intensive and lengthy processes that can span months.
Due to the high demand for domain expertise, existing automatic summarization methods fail to accurately identify relevant studies and generate high-quality summaries.
To that end, we introduce \sysname{}, a human-centered interactive AI agent powered by large language models that revolutionizes the systematic review workflow.
\sysname{} partitions a large literature corpus based on semantics and employs a multi-agent design for more focused processing of literature, leading to significant improvement in the quality of generated SRs.
\sysname{} also provides intuitive visualizations of the corpus and agent trajectories, allowing users to effortlessly monitor the actions of the agent and provide real-time feedback based on their expertise.
Our user studies with 9 medical professionals demonstrate that the visualization and interaction mechanisms can effectively improve the quality of synthesized SRs by 27.2\%, reaching 79.7\% of human-written quality. At the same time, user satisfaction is improved by 34.4\%. 
With \sysname{}, it only takes a clinician about 1.5 hours, rather than months, to complete a high-quality systematic review.
\sysname{} demonstrates great potential in facilitating more timely and informed decision-making in high stake application scenarios\footnote{Code and data are available at: \url{https://github.com/OSU-NLP-Group/InsightAgent}.}.
\end{abstract}

\section{Introduction}

Systematic reviews (SRs) are the cornerstone of evidence-based practice (EBP) across high stakes disciplines, such as healthcare, providing comprehensive synthesis of research evidence to inform clinical decision-making \cite{EBP}. 
Notably, the number of published SRs indexed in PubMed per year has increased from less than 50 in the 1990s to almost 36,000 in 2022 \cite{brignardello2025systematic}, indicating huge amount of human effort have been dedicated to conducting SRs.
Nevertheless, conducting systematic reviews remains a labor-intensive and time-consuming process that can take months to complete \cite{chandler2019cochrane, SR-bottleneck}. 

The systematic review process comprises several key steps: 
formulating a research question, collecting a corpus of literature, defining inclusion and exclusion criteria, 
screening relevant records, summarizing these studies, synthesizing the findings, and generating a final report.
While the initial steps are well-supported by information retrieval tools, the latter stages — specifically, record screening, literature summarization, and finding synthesis — remain significant bottlenecks \cite{SR-bottleneck}.
These steps demand intensive effort in reading, comprehending, and integrating information from a large volume of studies, posing challenges in efficiency and consistency.

So far, automation techniques for SRs mainly focus on record screening and shows varying sensitivity (recall), resulting in only limited adoption and modest saving of manual labor \cite{toth2024automation}.
For the evidence synthesis stage, more recent literature review systems based on large language models, such as ChatCite \cite{chatcite} and AutoSurvey \cite{autosurvey}, still overlook subtle but important details, resulting in generic summaries with untraceable sources, making them unsuitable for rigorous systematic reviews.

\begin{figure*}[ht]
    \centering
    \includegraphics[width=\linewidth]{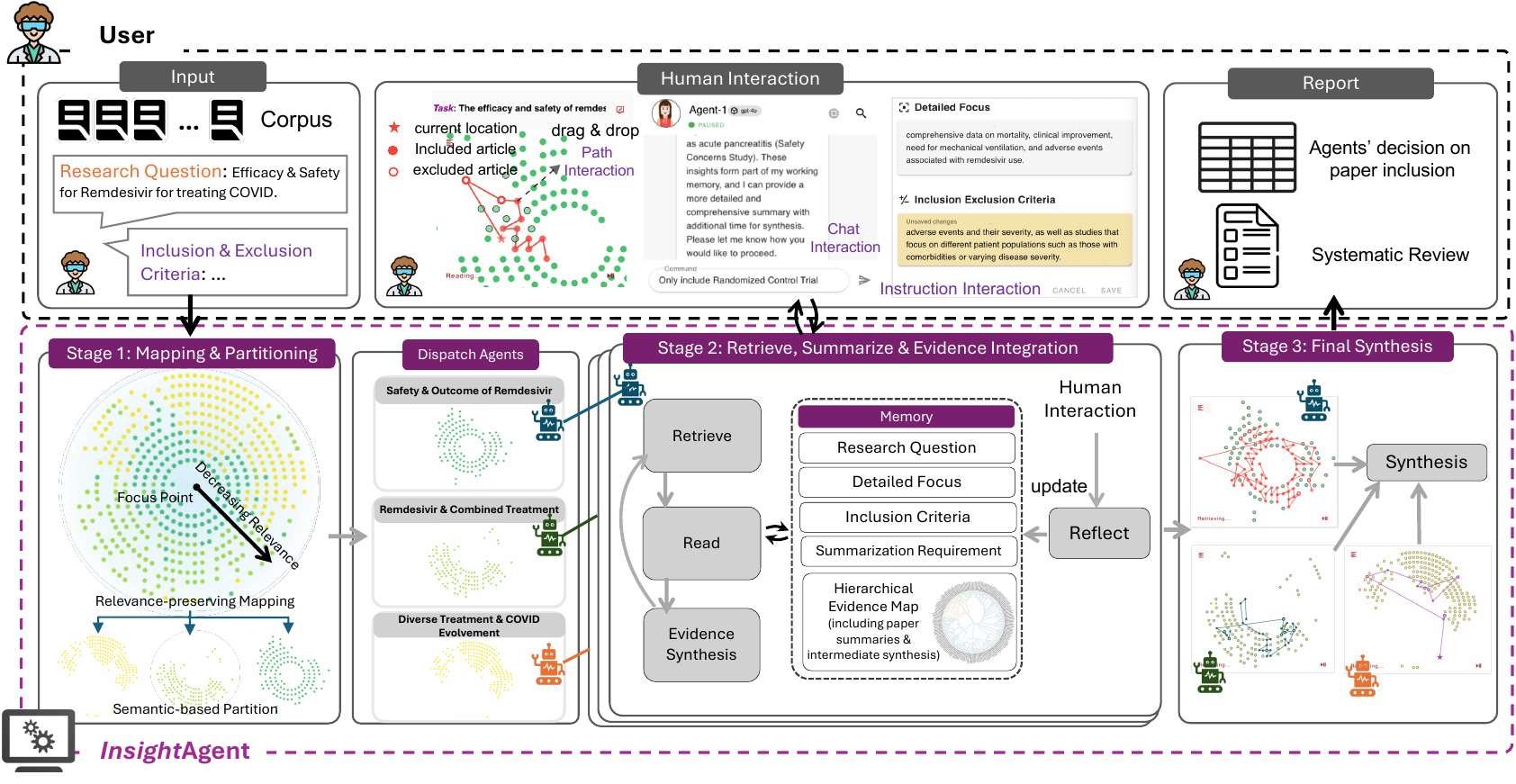}
    \caption{Overview of the \sysname{} workflow. In Stage 1, the corpus is mapped into semantic clusters. In Stage 2, multiple agents concurrently read and synthesize evidence under real‐time user guidance for each cluster. Finally, in Stage 3, findings of all agents are integrated into a complete systematic review.}
    \label{fig:system-overview}
\end{figure*}

To overcome these limitations, we introduce \textbf{\sysname{}}, the first automated system capable of generating high-quality systematic reviews. We propose a human-centered agent framework that equips language agents \citep{language-agent-tutorial} with a user-friendly graphical interface which enables real-time user oversight using visualization and the incorporation of user expertise via interactions. 
As illustrated in Figure \ref{fig:system-overview}, \sysname{} projects a large corpus into a circular \textit{relevance-preserving map} \cite{RPM}, where more relevant articles appear toward the center and semantically similar articles are clustered together, and dispatches multiple agents to read and summarize these clusters in parallel. This multi-agent design is inspired by the recommended multi-reviewer strategies in systematic reviews \cite{chandler2019cochrane}, where assigning different subsets of studies to independent reviewers can reduce individual bias and accelerate the initial screening. Each agent then explores the cluster from the center, identifies and reads relevant articles, and integrates findings into an interim SR. In this process, \sysname{} simultaneously generates a provenance tree that clearly tracks supporting articles for each interim finding.

With these visualization techniques, users can intuitively monitor the reading trajectory of each agent and intervene via various types of interactions to adjust the focus of the agent when needed.
At last, users can inspect the provenance tree to ensure the summaries in the output SR are properly supported by evidences.

We comprehensively evaluate the effectiveness of \sysname{} using 15 existing systematic reviews in the biomedical domain with 9 domain experts and medical students.
With GPT-4o \cite{openai2024gpt4ocard} as the backbone, the multi-agent design of \sysname{} allows it  to produce systematic reviews with 15.6\% higher quality than AutoSurvey.
Interactions with users further improve article identification accuracy by 47\% (F1 points), generated review quality by 27.2\%, and overall user satisfaction by 34.4\%.
On average, a clinician can complete a systematic review in 1.5 hours using \sysname{} and reach 79.7\% of human-written quality.
\sysname{} drastically cuts the time needed for crafting a high-quality systematic review from months to hours, demonstrating the great potential of human-centered AI agents in accelerating evidence-based discoveries.

\section{Related Work}

\subsection{Language} Agents for Literature Survey
Recent advancements in large language models (LLMs) have opened new possibilities for automate the literature review process.
Several works have applied LLMs to automate literature reviews. AutoSurvey~\cite{autosurvey} generates literature summaries by constructing an outline and progressively refining it, while ChatCite~\cite{chatcite} extracts key elements from research papers and incrementally generates task-specific summaries. LitLLM~\cite{litllm} retrieves papers through keyword-based queries and produces summaries using zero-shot generation methods. Following a similar paradigm, \citet{instruct} take a step-by-step approach and generate sections of a literature survey in sequence; \citet{pathfinder} facilitate semantic exploration of astronomical literature using LLMs to improve context-based retrieval. 

So far, existing LLM agents for literature review mostly operate in a fully autonomous fashion.
The lack of user interaction and transparency in these systems presents significant limitations.
Autonomous agents without human involvement often struggle to maintain coherence and transparency in their decision-making processes.
Our proposed system, \sysname{}, addresses these gaps by enabling real-time user monitoring and intervention for the agents' decision making through an intuitive graphical user interface. 
Through a human-centered interface, users can visually monitor agents’ tasks, guide their progress, and interact with them to ensure coherence and relevance.

\subsection{Visual Analytics for Information-seeking and Decision-making}
Visual analytics (VA) methods embed visualization into the data analysis processes and can effectively facilitate decision-making and information-seeking \cite{ isenberg2016visualization, lee2020cerc, qiu2022docflow}. 
In the context of information-seeking, VA has been applied primarily in two ways: (1) sense making and interpretability, and (2) retrieval, classification, and decision-making.

\paragraph{Sensemaking and Interpretability.}
VA systems assist researchers in comprehending thematic and relational structures within extensive document collections. For instance, HINTs \cite{lee2024hints} employ hypergraph representations to highlight complex entity-topic relationships, whereas \citet{RPM} utilize adaptive 2D layouts to map documents according to user queries.

\paragraph{Retrieval, Classification, and Decision-Making.}
VA methodologies also focus on targeted tasks like document retrieval and classification, which are crucial in systematic reviews.
Docflow \cite{qiu2022docflow} categorizes documents in response to user-specified queries through answer embedding similarity to streamline the record screening process. Studies also suggest that coupling machine learning–based retrieval with interactive visualization can significantly improve precision and recall in document retrieval and information-seeking\cite{da2023evaluating}. Beyond retrieval, research has shown that thoughtful interface design reduces cognitive biases \cite{cho2017anchoring,oral2023information} and facilitates strategic planning \cite{nazemi2022visual}. 

Building on these insights, our approach leverages \emph{LLM-driven agents} with a spatial document layout to facilitate systematic reviews, from where users can observe agent actions, refine corpus exploration, and achieve more effective evidence synthesis through a transparent, VA-based interface.
\section{\sysname{}}

In this section, we present \sysname{}, a human-centered approach that equips autonomous language agents with an interactive graphical user interface for improved summary quality and user satisfaction in high stakes domains.
\sysname{} operates in three stages: (1) \textbf{corpus mapping and partitioning}, (2) \textbf{record screening and evidence synthesis}, and (3) \textbf{final synthesis}.
In each stage, \sysname{} harnesses the capabilities of LLMs and leverage domain knwoledge from expert users, ensuring that the systematic review process is efficient and accurate.

\subsection{Stage 1: Corpus Mapping \& Partitioning}
\label{sec:mapping-partitioning}

The first stage of \sysname{} aims to project a large biomedical corpus into an intuitive layout and partition it for parallel processing by multiple agents. This mapping not only prepares a more focused action space for each agent, but also provides a visual interface for the user to monitor agent trajectories and intervene when needed  (Figure~\ref{fig:system-overview}). 

\paragraph{Corpus Mapping.}
To visualize the overall structure of the corpus, we use the radial-based relevance and similarity map (RSS map) \cite{RPM}.
Each article in the corpus is presented as a dot, whose positions is decided by two factors: (1) relevance to the \textit{research question} - more relevant articles appear closer to the center, and (2) semantic similarity to other articles - semantically similar articles are placed in a nearby region.

\paragraph{Corpus Partitioning.}

Once the documents are positioned in the radial layout, we apply K-means clustering to partition the corpus into semantically distinct clusters. The optimal number of clusters $k$ is automatically determined by the Elbow method~\cite{onumanyi2022autoelbow}, which evaluates cluster compactness through within-cluster and inter-cluster distances, resulting in an average of nine clusters across our evaluations. Subsequently, multiple agents are instantiated and operate in parallel, with each assigned to a separate cluster and an optional reading focus. Users may flexibly refine these clusters and adjust agents' reading based on their domain knowledge and research objectives. This corpus partitioning reduces noise and individual agent workload, significantly enhancing retrieval accuracy and summarization quality. A quantitative comparison demonstrating the effectiveness of our multi-agent design over a single-agent approach is presented in Appendix~\ref{appendix-evaluation-k}.

\subsection{Stage 2: Reading and Evidence Synthesis}
\label{sec:retrieval-summarization-evidence}

In this stage, each agent explores its assigned cluster, identifies relevant documents, generates incremental summaries, and integrates newly acquired evidence into their knowledge base. Throughout these processes, \sysname{} offers rich visualization and interactive controls so that users can easily monitor and refine the agents’ reading strategies to enhance the quality of the final systematic review (Figure~\ref{fig:system-overview}-Step 2).

\paragraph{Agent Setup and Record Screening.} When an agent is created for a given corpus, it is initialized with a \textit{research question} $Q_i$, \textit{inclusion and exclusion criteria} (e.g., study type), and a \textit{summary requirement} (e.g., desired level of detail or a specific analytic focus). The agent then begins screening articles from the center of the RSS map—corresponding to the most relevant documents—and progressively moves outward. At each step, the agent selects its next article from a defined \emph{receptive field}, consisting of the immediate neighbors of the current document within the relevance-preserving map. To maintain consistency across all agent operations, we standardize this receptive field to always include eight nearest neighboring articles. This design ensures agents systematically navigate through the corpus, reducing randomness and promoting coherent exploration. As each article is reviewed, the agent determines whether it satisfies the predefined inclusion criteria, dynamically updating its short-term reading strategy accordingly.

\paragraph{Summary Generation \& Memory Mechanism.}
For each relevant article, the agent summarizes the key findings and their relevance to $Q_i$. These summaries are stored in a \textit{local memory}, along with metadata such as timestamps and article embeddings.
Whenever the agent encounters overlapping or contradictory information, it merges existing summaries with the new one:
\begin{equation}
    M_{k+1} = f(M_k, S_j),
\end{equation}
where $S_j$ is the freshly generated summary for the current document, and $M_k$ is a previously stored interim evidence synthesis.
This incremental evidence synthesis process avoids redundancy and gradually constructs a coherent subdomain knowledge base.
Importantly, each agent’s memory remains isolated from others until the final synthesis stage, allowing it to develop a specialized perspective on its assigned topic.

\paragraph{Transparent Evidence Integration.}
To maintain accountability for how each conclusion is formed, \sysname{} logs every summary merge and evidence update in a \textit{dependency graph}, which functions as an attribution or provenance structure.
Leaf nodes in this tree represent article-level summaries and other nodes represent interim syntheses.
Different color-coding or labeling denotes contributions from distinct agents, allowing domain experts to verify sources and scrutinize any disputed findings.

\paragraph{User Interventions.}
\sysname{} offers three types of real-time user interaction interfaces to effectively collaborate with users in the systematic review process:
\begin{itemize}
    \item \textbf{Path Navigation.} On the RSS map, users can \textit{drag} an agent’s pointer to a missed relevant article and the agent will read the article next and update its local memory accordingly.
    \item \textbf{Chat Navigation.} Users can issue natural-language directives (e.g., ``Focus on randomized controlled trials’’), causing the agent to reflect on changes to $Q_i$, inclusion criteria, or summary requirements. It then revises its retrieval strategy and merges new findings or discards outdated ones.
    \item \textbf{Instruct Navigation.} For more fine-grained control, experts can directly edit the agent’s parameters, such as specifying stricter inclusion criteria or change to a different summarization format. Upon receiving these instructions, the agent double-checks relevant memory entries to ensure previously stored summaries align with the updated requirements.
\end{itemize}
Whenever user interventions alter the agent’s behavior, the agent enters a \textit{reflection phase}, during which it reconciles any conflicts in local memory, and adjusts its reading strategy to be consistent with the latest directives.
By combining iterative retrieval, localized summarization, and user-driven oversight, \sysname{} builds a flexible, transparent evidence base that will later feed into the final synthesis stage. We report the details of our interface design in appendix~\ref{fig:insightagent-ui}.

\subsection{Stage 3: Final Synthesis}
\label{sec:final-synthesis}

Once the dispatched agents have finished building localized evidence bases, \sysname{} integrates these subdomain findings into a coherent final summary following the template specifies by the user.
The template usually has several sections, including 
\emph{Introduction}, \emph{Study Design}, \emph{Key Findings}, \emph{Discussion}, and \emph{Conclusion}. Citations in the resulting extended abstract follow a structured format: \sysname{} uses \texttt{[citation\_number]} to refer back to original documents or interim summaries, which ensures the supporting sources for each argument are clearly traceable.
The dependency tree is updated accordingly. We provide an example final synthesis template in Appendix \ref{appendix-prompt-template}.

\section{Experiments}
\label{sec:eval-setup}

\subsection{Experiment Setup}

We perform a comprehensive human evaluation to assess \sysname{}'s effectiveness in accelerating systematic reviews. Our study involves 15 published systematic reviews.
Among them, 13 were published in 2024 and 2 were from 2022 and 2023.
We reconstruct the literature corpus following their published search strategy, with corpus sizes ranging from 72 to 7{,}356 articles (the average inclusion rate is 5.7\%).
Details of the chosen systematic reviews are provided in Appendix \ref{appendix: sr-datasets}.
Following common practices in record screening and considering the cost and LLM context length constraints, we only use the title and abstract of each article in this study.
We invite 9 medical experts with prior experience in publishing systematic reviews to participate in the evaluation.

Our study primarily focuses on four axes:

\begin{enumerate}[leftmargin=1em,label=(\arabic*)]
    \item \textbf{Record Screening.}
    We measure how accurately can \sysname{} identify relevant studies in the corpus during record screening. We report precision, recall and F1 scores.

    \item \textbf{
    Systematic Review Quality.
    }
    We use a rubric-based method in which each generated report is independently scored by two experts on multiple dimensions, such as comprehensiveness and writing quality, totaling 100 points. 
    Each systematic review is scored by two different experts and we report the average rating.
    The rubric is collaboratively defined by domain experts, with peer-reviewed human systematic reviews as the ground truth (100 points).  
    Appendix \ref{appendix-evaluation-rubric} provides the detailed rubric.
    \item \textbf{
    User Experience
    } 
    After using each system, participants complete a questionnaire evaluating the usability of the interface, perceived precision, and overall satisfaction. We also conduct brief follow-up interviews for qualitative feedback on transparency, user control, and confidence in the output systematic review. The questionnaire is available in Appendix \ref{appendix-usability-questionnair}.

\end{enumerate}

\paragraph{Evaluation Procedure.} Each of the nine experts is randomly assigned two to four systematic reviews, with each systematic review independently evaluated by at least two experts familiar with its topic. To ensure fairness and consistency in comparing different methods (\sysname{}, AutoSurvey, and ChatCite), we assign reviews generated by different methods for the same systematic review to the same expert.

While we do not set a time limit, we observe that on average it only takes participants about 1.5 hours to finish a session and produce a systematic review. This marks a substantial speedup compared to manual systematic reviews, which typically takes months.

\paragraph{Implementation Details.}
We experiment with two popular large language models as the system backbone: the proprietary GPT-4o \cite{openai2024gpt4ocard} and the open-weight Llama 3.3 70B \cite{grattafiori2024llama3herdmodels}.
We use default hyperparameters for text generation. 

To evaluate the effectiveness of user interactions, we test two variants: \sysname{}, the default configuration that supports real-time user interactions, and \sysname{}$_{auto}$, which disables interactions and autonomously completes a systematic review in a few minutes.

\paragraph{Baseline Systems.}
We compare \sysname{} with two recent fully autonomous LLM-based literature review systems:
\begin{itemize}[leftmargin=1em]
    \item \textbf{ChatCite}~\cite{chatcite}, an incremental reflective summarization system for literature review.
    \item \textbf{AutoSurvey}~\cite{autosurvey}, an LLM-based system combining embedding-based retrieval with automated survey generation.
\end{itemize}

For ChatCite, we follow \citet{chatcite} and implement the key element extractor and comparative generator in biomedical settings.
For AutoSurvey, we use the released implementation and retrieve the top 100 articles for each SR for summarization.

\begin{figure*}[ht!]
    \centering
     \includegraphics[width=\linewidth]{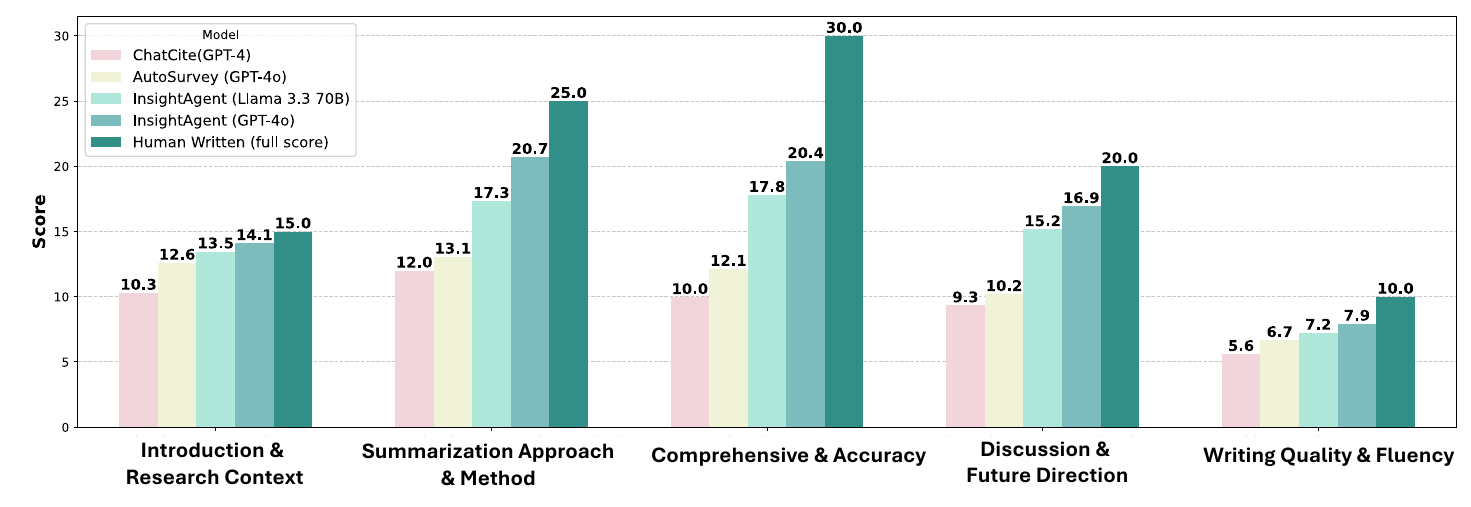}
    \caption{Detailed comparison of summarization quality of \sysname{} against ChatCite and AutoSurvey across five evaluation dimensions: Introduction \& Research Context, Summarization Approach \& Method, Comprehensiveness \& Accuracy, Discussion \& Future Directions, and Writing Quality \& Fluency.}
    \label{fig:summary-quality-comparison-detailed}
\end{figure*}

\subsection{Experiment Results}
\subsubsection{Record Screening}

\begin{table}[ht!]
\centering
\small
\setlength{\tabcolsep}{0pt}
\renewcommand{\arraystretch}{1.0}
\begin{tabular}{lccc}
\toprule
\textbf{System} & \textbf{Recall (\%)} & \textbf{Precision (\%)} & \textbf{F1\%}\\ 
\midrule
BM25 (Top-100) & 54.30 & 16.50 & 25.3 \\
ChatCite & -- & -- & --\\
AutoSurvey (Top-100) & 70.30 & 20.43 & 31.6  \\
\midrule
\sysname{}$_{auto}$ (\text{Llama 3.3}) & 66.40 & 62.40 & 64.3 \\
\sysname{} (Llama 3.3) & \underline{87.90} & \textbf{80.10} & \underline{83.8} \\
\sysname{}$_{auto}$ (\text{GPT-4o}) & 71.10 & 51.90 & 60.0 \\
\sysname{} (GPT-4o) & \textbf{98.50} & \underline{79.80} & \textbf{88.2} \\
\bottomrule
\end{tabular}
\caption{Performance in record screening. We bold the best performance and underline the second best.}
\label{tab:relevant_articles}
\end{table}

Table \ref{tab:relevant_articles} presents the performance of each system in identifying relevant articles. We include BM25 \cite{robertson1994bm25} for reference.
We restrict both BM25 and AutoSurvey's retrieval to top-100, which is a practical cutoff given that all of the systematic reviews in our evaluation include fewer than 100 articles.
ChatCite does not perform retrieval, instead it summarizes all user-provided articles. Hence, we omit it from this comparison.

The results indicates that \sysname{}$_{auto}$ with both GPT-4o and Llama 3.3 outperform BM25 and AutoSurvey (GPT-4o) by a large margin in record screening, with a substantial advantage in precision (62.4\%/41.9\% vs 20.4\%), demonstrating the effectiveness of our multi-agent design and corpus partitioning strategy in more accurately identify articles relevant to the research question.

Impressively \sysname{} further substantially improves both recall and precision, with \sysname{}(GPT-4o) reaching a near-perfect 98.5\% recall.
These results show that our user-centered design with an interactive interface can effectively help users monitor the agent's reading progress and correct agent mistakes based on their domain knowledge. 
The comprehensive and accurate record screening results lays a solid foundation for \sysname{} to generate high-quality systematic reviews in the final synthesis stage.

\subsubsection{Quality of the Generated Summaries}

\begin{table}[ht]
  \centering
  \small
  \setlength{\tabcolsep}{4pt}
  \begin{tabular}{@{}l c@{}}
    \toprule
    \textbf{System} & \textbf{Score} \\ 
    \midrule
    ChatCite (\text{GPT-4}) & 47.1 \\
    AutoSurvey (\text{GPT-4o}) & 54.0 \\
    \sysname{}$_{auto}$ (\text{Llama 3.3}) & 60.9 \\
    \sysname{} (\text{Llama 3.3}) & \underline{70.2} \\
    \sysname{}$_{auto}$ (\text{GPT-4o}) & 62.4 \\
    \sysname{} (\text{GPT-4o}) & \textbf{79.7} \\
    \bottomrule
  \end{tabular}
  \caption{Quality of generated systematic reviews rated by human experts. We bold the best performance and underline the second best.}
\label{tab:summary_quality}
\end{table}

Table~\ref{tab:summary_quality} presents the quality of generated systematic reviews evaluated by experts.
We observe that \sysname{}$_{auto}$ using the weaker Llama 3.3 model already outperform both baselines using more powerful GPT-4 and GPT-4o models.
Compared to AutoSurvey, \sysname{}$_{auto}$(GPT-4o) improves generated review quality by 8.4 points, marking a significant step forward for fully autonomous agents for systematic reviews.

When human oversight and guidance is available, \sysname{} with both base LLMs further show substantial improvement in review quality by 9.3 points (Llama 3.3) and 17.3 points (GPT-4o).
\sysname{}(GPT-4o) reaches a quality rating of 79.7 points, making it the first system to be practically useful for domain experts to accelerate the systematic review process.

\begin{figure}[ht]
    \centering
    \includegraphics[width=\linewidth]{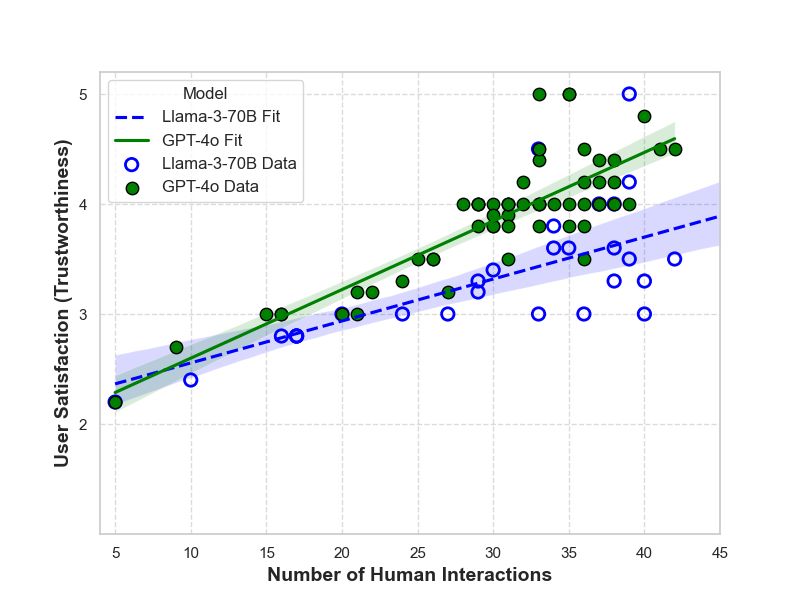}
    \caption{Relationship between the number of human interactions and perceived system trustworthiness for GPT-4o and LLamA 3.3 70B. Scatter points represent individual data samples, while regression lines with confidence bands illustrate the overall trend. }
    \label{fig:correlation-interaction-trust}
\end{figure}

\begin{table*}[ht]
\centering
\small
\renewcommand{\arraystretch}{1} 
\begin{tabular}{p{0.3\linewidth}cccc}
\toprule
\multirow{2}{10cm}{\textbf{Question}} & \multirow{2}{2cm}{\centering\textbf{\sysname{}$_{auto}$ (Llama 3.3)}} &  \multirow{2}{2cm}{\centering\textbf{\sysname{} (Llama 3.3)}} & \multirow{2}{2cm}{\centering\textbf{\sysname{}$_{auto}$ (GPT-4o)}}  &  \multirow{2}{2cm}{\centering\textbf{\sysname{} (GPT-4o)}} \\
\\
\midrule
System was easy to use.        
& 2.0/5 & \underline{3.9}/5 & 2.1/5 & \textbf{4.0}/5 \\
Confidence in recommendations. 
& 2.6/5 & \underline{4.2}/5 & 2.8/5 & \textbf{4.5}/5 \\
Visualizations aided understanding. 
& 2.9/5 & \textbf{4.2}/5 & 2.9/5 & \textbf{4.2}/5 \\
Ability to guide or correct agents. 
& 2.7/5 & \underline{3.9}/5 & 2.9/5 & \textbf{4.6}/5 \\
Overall satisfaction.          
& 3.0/5 & \underline{4.0}/5 & 3.2/5 & \textbf{4.3}/5 \\
\bottomrule
\end{tabular}
\caption{Usability \& Trustworthiness Likert Scores. We bold the best performance and underline the second best.}
\label{tab:usability}
\end{table*}

We break down the quality improvement by evaluation dimensions in Figure \ref{fig:summary-quality-comparison-detailed}. While \sysname{} shows improvement in all aspects, the advantage more prominent in the comprehensiveness and accuracy of reviews and the derived insights and conclusions. 
Qualitatively, evaluators find that \sysname{} consistently produces more comprehensive and relevant summaries than ChatCite and AutoSurvey, which frequently include irrelevant information partly due to their low precision in record screening. Furthermore, \sysname{} is able to deliver more reliable findings, sometimes even suggesting novel insights absent in the original human-written SR.
At last, participants report that \sysname{}'s graphical interface and well-designed interaction features make it easier to trace evidence, significantly boosting their confidence in the generated reviews.
We present a detailed qualitative comparison for one SR in Appendix \ref{appendix-reports-with-evaluation}. 

\begin{figure*}[ht]
    \centering
    \includegraphics[width=\linewidth]{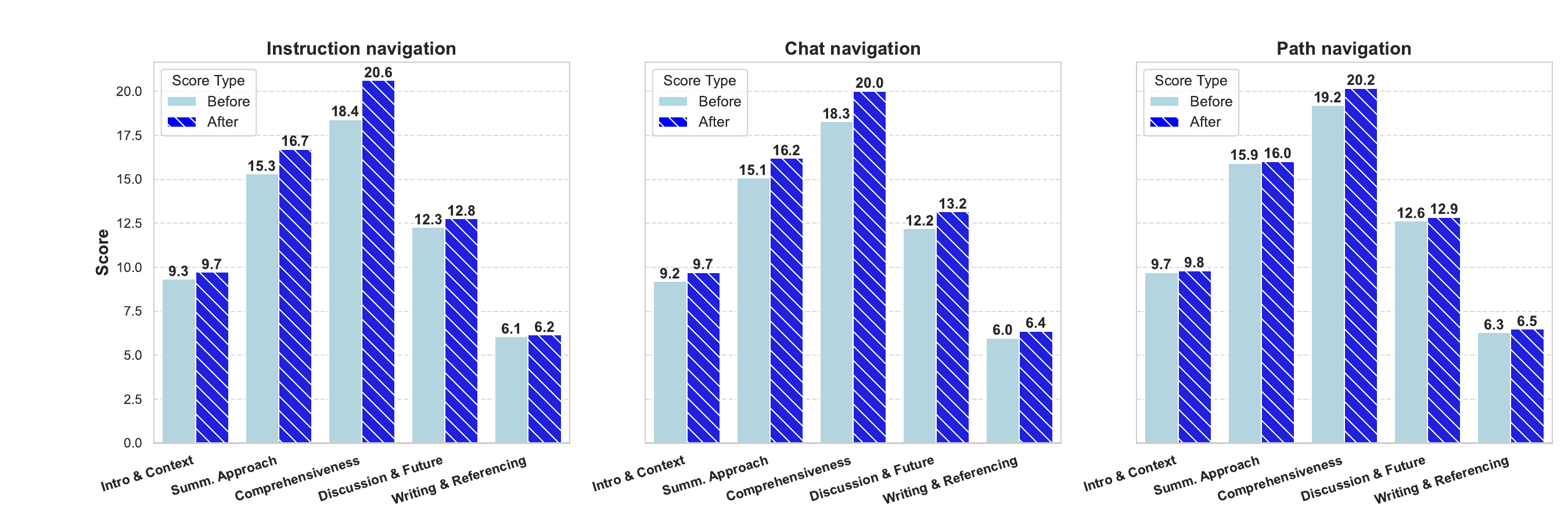}
    \vspace{-2em}
    \caption{Impact of interaction types on summary quality improvement across different evaluation perspectives. Each subplot compares scores before and after interaction for Instruction, Chat, and Path navigation.}
    \label{fig:interaction-effectiveness-detail}
\end{figure*}

\subsubsection{User Experience}
After each user study session, we administer a questionnaire to collect feedback on the usability and trustworthiness of \sysname{}, shown in Table \ref{tab:usability}. Participants evaluated \sysname{} in both \textit{autonomous} and \textit{interactive} modes. While \sysname{} with GPT-4o generally achieves higher scores (e.g., 4.4/5 for “System was easy to use” compared to 3.9/5 for Llama 3.3 70B), both backbone models can benefit from user guidance.
For instance, the \textit{confidence in recommendations} metric increases from 2.8/5 to 4.2/5 for GPT-4o and from 2.6/5 to 4.0/5 for Llama 3.3 70B, indicating that real-time oversight and domain-expert input significantly boost trust in the system’s output. 

To examine how interactions influence the user experience of \sysname{}, we plot the user satisfaction scores and the number of interventions for each SR in Figure \ref{fig:correlation-interaction-trust}. Overall, both GPT-4o and Llama 3.3 70B exhibit a clear trend: as participants engage in more interventions, their confidence in the system consistently grows. Among the two backbone LLMs, GPT-4o demonstrates stronger abilities in collaborating with users, echoing our results in record screening and final synthesis stages. 

These quantitative results clearly demonstrate the practical benefits of incorporating human interaction into \sysname{}. A paired t-test confirms that the observed improvements are statistically significant: for \sysname{} (GPT-4o), interaction increases summary quality by 27.2\% ($p=3.43\times10^{-7}$) and user satisfaction by 34.4\% ($p=1.89\times10^{-6}$). These statistically robust improvements underline the substantial impact of human guidance within the review process 

These quantitative findings are further supported by qualitative feedback gathered from post-session interviews. Participants repeatedly highlight the importance of being able to \textit{direct} the agent’s reading paths and \textit{monitor} its exploration dynamics through visualizations:
\begin{itemize}[leftmargin=1.2em]
\item \textit{“I felt more in control when I could redirect the agent to areas I knew were important.”}

\item \textit{“Visualizing exploration paths helped me trust that critical topics weren’t being missed.”}
\end{itemize}

Overall, these results show that \sysname{}’s interactive features successfully enhance user confidence through engagement. In the following section, we further show  how different forms of user intervention contribute to these improvement.

\subsubsection{Effectiveness of Interactions}

\sysname{} supports diverse types of interactions, offering users flexible control in the systematic review process. 
To rigorously evaluate the impact of a single user interaction, we use \sysname{}$_{auto}$ to complete two interim synthesis before and after an interaction.
For each of the three interaction types, we sample 50 such interim synthesis pairs in our user studies and ask participants to score the auto-completed systematic reviews. The results are reported in Figure \ref{fig:interaction-effectiveness-detail}.

\paragraph{Instruction Navigation.}
By directly revising the research question, inclusion \& exclusion criteria, or summarization requirements, users effectively realign the agent’s entire reading and synthesis strategy.
As illustrated in \figurename~\ref{fig:interaction-effectiveness-detail} (left),
instruction navigation noticeably improves the coverage of relevant literature and findings in the synthesized SRs.
Although participants less frequently use this type of interaction (see Appendix~\ref{appendix-interaction-stats} for usage statistics), they consistently cite it as a powerful way to ``reset'' or ``refine'' the agent’s focus, thus yielding pronounced benefits in final reporting quality.

\paragraph{Chat Navigation.}
As shown in \figurename~\ref{fig:interaction-effectiveness-detail} (center), chat navigation brings modest improvement across all dimensions.
We find chat is used in a flexible way. Some participants use chat primarily for asking clarifying questions without intervening the agent's decision-making, leading to minimal changes in the final summary.
In other cases, users leverage chat to propose new angles of investigation or update the summarization format, significantly improving the completeness of the review.

\paragraph{Path Navigation.}
Figure \ref{fig:interaction-effectiveness-detail} (right) shows that path navigation also exhibits moderate but consistent improvements, especially in the comprehensiveness of SRs. 
By pinpointing overlooked articles in the corpus, participants can ensure that relevant studies were included, thereby enhancing the coverage of the final review.

\subsection{Error Analysis}
While \sysname{} has made a significant stride toward automating systematic reviews, we still identify two key limitations upon comparing agent generated SRs with human-written ones (See Appendix \ref{appendix-reports-with-evaluation} for detailed SRs, evaluations, and analysis). A more detailed error analysis is in Appendix~\ref{appendix-detailed-error-analysis}.

\paragraph{Insufficient Statistical Analysis.}
Human experts often perform statistical analysis over data from multiple relevant studies to derive numeric evidence that supports rigorous conclusions.
While \sysname{} shows promise in evidence synthesis, it is not yet capable of performing such analysis.

\paragraph{Limited Planning and Evidence Weighting.}
Human-generated systematic reviews often control the proportion of articles drawn from different sources, such as high-quality randomized controlled trials (RCTs) versus observational studies (e.g., “60\% of data is from RCTs and 40\% from observational cohorts”) and weight these sources accordingly.
While \sysname{} can accurately identify relevant studies, it generally treats the articles equally and has limited capacity in considering such global constraints and adjust review plans.

These limitations call for future research in augmenting AI agents systematic reviews with advanced analysis and planning modules. 
\section{Conclusion}
We introduced \sysname{}, a human-centered language agent for accelerating systematic reviews.
\sysname{} adopts a novel multi-agent design and is equipped with an intuitive graphical interface that supports both agent decision monitoring and user interactions, resulting in improved quality and user experience in high stakes domains like healthcare.
Through comprehensive human study, we show that a single domain expert can finish a high-quality systematic review in only 1.5 hours, rather than months, and reach 79.7\% of human-written quality.
Our work demonstrates the great potential of interactive AI agents in accelerating systematic reviews and further facilitate scientific research.
\section*{Limitations}
Despite promising results, this work faces three primary limitations. 
First, we conducted a relatively small-scale user study, owing to both the high human effort required for evaluation and the limited availability of fully completed systematic reviews. 
Second, due to cost and LLM context length limits, this work's setup is restricted to reading only the abstract of each study rather than full texts, potentially omitting critical details
that can shape final conclusions. 
Third, the system lacks the ability to extract and synthesize numerical statistics or weight evidence based on study design; consequently, effect sizes, incidence rates, and other quantitative measures are not rigorously reported, and stronger studies are not afforded greater influence in the synthesized results. We deem developing more advanced systematic review agents capable of quantitative analysis and evidence weighting as promising and important future work directions.

\bibliography{custom}
\pagebreak
\appendix
\section{Prompt Template}

\label{appendix-prompt-template}
we present the complete prompt in \sysname{} for retrieve (Prompt~\ref{prompt-retrieves}), read (Prompt~\ref{prompt-read}), synthesis (Prompt~\ref{prompt-synthesis}), and reflection (Prompt~\ref{prompt-reflect}) actions.

\section{Quantitative Evaluation of Multi-Agent Design}
\label{appendix-evaluation-k}

To quantitatively demonstrate the effectiveness of the multi-agent design, we compare the retrieval performance of InsightAgent under two different conditions: an automatically determined number of clusters (multi-agent scenario) and a single-cluster scenario (single-agent). The evaluation metrics considered include recall, precision, and the F1 score. Results are summarized in Table~\ref{tab:multi-agent-comparison}, which indicates that the multi-agent configuration, using an automatically determined number of clusters (averaging nine), consistently outperforms the single-agent scenario across both evaluated LLM backbones (Llama 3.3 and GPT-4o). Specifically, the multi-agent setup yields notably higher recall, precision, and F1 scores, underscoring its superior capability to identify relevant articles accurately. In contrast, the single-agent design results in a substantial performance decline, highlighting the importance of corpus partitioning and parallel processing for ensuring the quality of review output.

\begin{table*}[ht]
\centering
\setlength{\tabcolsep}{8pt}
\renewcommand{\arraystretch}{1.2}
\caption{Retrieval Performance Comparison Between Single-Agent and Multi-Agent Designs}
\label{tab:multi-agent-comparison}
\begin{tabular}{lccc}
\toprule
\textbf{System Configuration} & \textbf{Recall (\%)} & \textbf{Precision (\%)} & \textbf{F1 (\%)} \\
\midrule
InsightAgent$_{auto}$ (Llama 3.3) - K$_{auto}$ (9 on avg.) & 66.4 & 62.4 & 64.3 \\
InsightAgent$_{auto}$ (Llama 3.3) - K = 1 & 56.6 & 52.0 & 54.2 \\
InsightAgent$_{auto}$ (GPT-4o) - K$_{auto}$ (9 on avg.) & 71.1 & 51.9 & 60.0 \\
InsightAgent$_{auto}$ (GPT-4o) - K = 1 & 60.9 & 44.7 & 51.6 \\
\bottomrule
\end{tabular}
\end{table*}

\section{\sysname{} Interface}
\label{system-interface}
\begin{figure*}[ht]
    \centering
    \includegraphics[width=\linewidth]{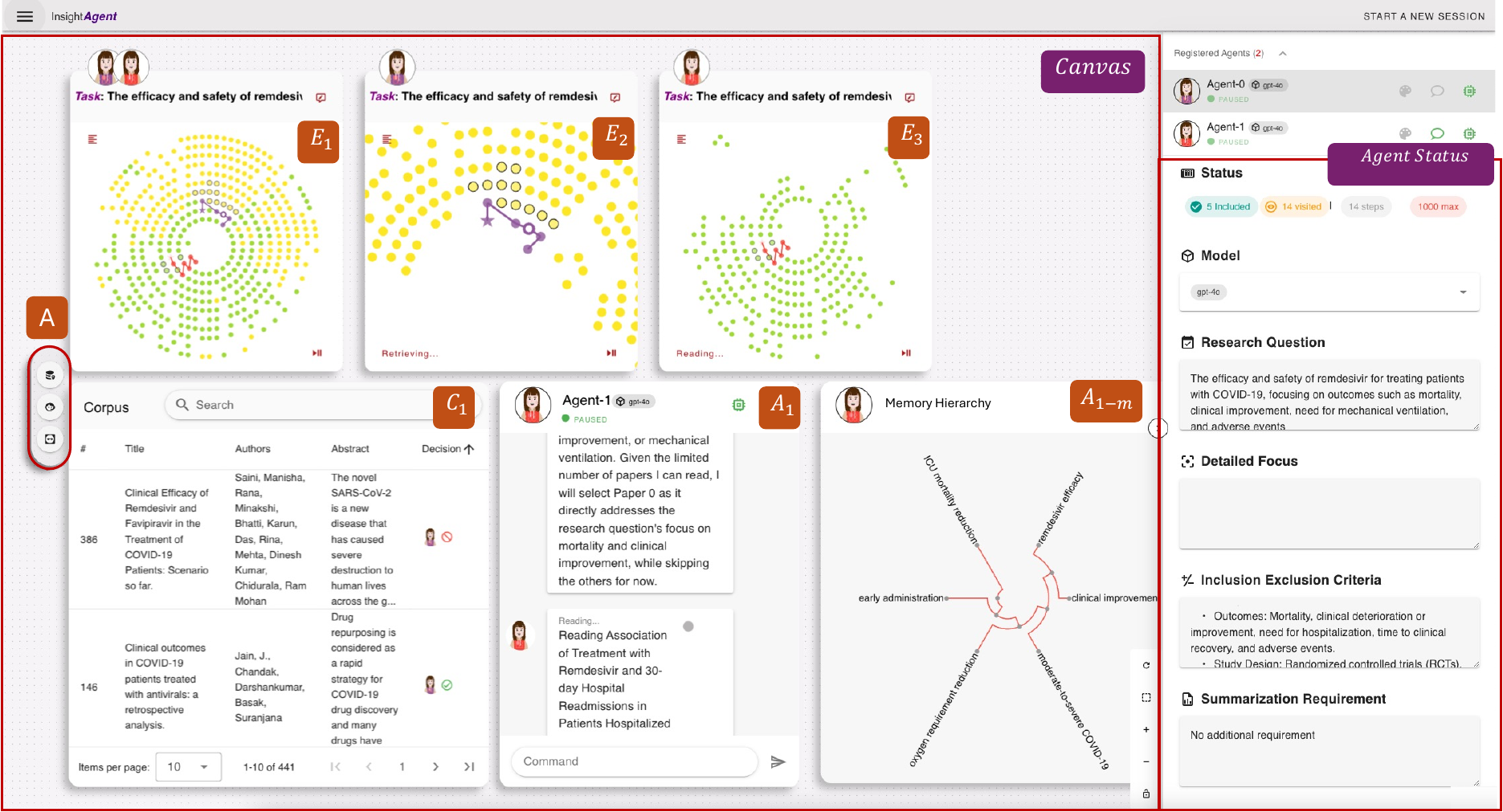}
    \caption{A screenshot of the \sysname{} interface while conducting a systematic review. The central \emph{Canvas} is an infinite- scrollable space for creating multiple \emph{Environments} (E1, E2, E3) and attaching any number of agents or collaboration panels. This design allows users to freely drag and drop agents, documents, or synthesis outputs as they refine the review process.
}
    \label{fig:insightagent-ui}
\end{figure*}

\sysname{} is designed to give users maximum flexibility in monitoring systematic reviews (Figure~\ref{fig:insightagent-ui}). The main \emph{Canvas} (\textbf{top center}) hosts an \emph{infinite scrollable area} where researchers can place one or more \emph{Environment} components (\emph{e.g.}, E1, E2, E3). Each environment projects a corpus of documents onto an interactive 2D map, displaying article distributions and agent ``line-of-action'' paths for reading. By hovering over any dot in an environment, users can view that article’s metadata (title, authors, abstract, etc.), then decide which agent should read or skip it. This canvas-based design affords more fluid iteration than a static layout, letting users freely reposition cards, open new environments, or consolidate findings as needed.

The \emph{Corpus View} (C1) displays metadata for each document. It tracks which agent has read a given paper, whether it was included or excluded, and can export those decisions in CSV form for reference or reporting. The \emph{Chat Window} (A1) offers a direct interface for conversations with a chosen agent, enabling chat-based instructions or updates (such as clarifying a misunderstanding or adjusting the focus of the agent’s reading). Alongside the chat, each agent’s \emph{Memory Hierarchy} (A1--M) visualizes how individual evidence items are synthesized into higher-level summaries. Users can expand or hover over nodes to review summarized findings and cross-check citations, further strengthening traceability.

The menu on the left (A) lists the system’s primary components: Environments, Agents, and a Collaboration Panel. Researchers can add multiple agents to a single environment—enabling parallel reading of different subtopics—or move a single agent across multiple environments by simply dragging its icon from one environment to another. The agent’s ongoing tasks and progress appear in the \emph{Agent Status} panel on the right, which also displays the research question, detailed focus, inclusion/exclusion criteria, and summarization requirements. These can be edited in real time, representing an \emph{instruction-based interaction} that updates the agent’s parameters mid-stream. Meanwhile, “path-based” interactions occur within the environment cards (E2, E3), where the user drags the agent pointer around the projected document space to direct its reading order. Finally, the Collaboration Panel allows multiple agents (Agent~0 and Agent~1, for example) to be grouped, letting them exchange findings or produce a unified synthesis through a shared chat or memory structure.

This flexible interface facilitates a more iterative, user-driven approach than is typical in static review tools. By combining path navigation, direct textual instructions, and real-time agent collaboration, \sysname{} broadens researchers’ capacity to organize, track, and refine systematic reviews in a manner best suited to their investigative goals.

\begin{table*}[ht]
  \centering
  \small
  \caption{Prevalence of Major Error Types in Extended Abstracts (number of reports flagged / total reports (\%)).}
  \label{tab:error-prevalence}
  \resizebox{\textwidth}{!}{%
    \begin{tabular}{
      l
      >{\centering\arraybackslash}p{3cm}
      >{\centering\arraybackslash}p{1.8cm}
      >{\centering\arraybackslash}p{1.8cm}
      >{\centering\arraybackslash}p{1.8cm}
    }
      \toprule
      \textbf{Error Type}
        & \multicolumn{1}{c}{\makecell{\textbf{ChatCite}\\($R=15$)}}
        & \multicolumn{1}{c}{\makecell{\textbf{AutoSurvey}\\($R=15$)}}
        & \multicolumn{1}{c}{\makecell{\textbf{InsightAgent}\\(Llama‑3.3, $R=30$)}}
        & \multicolumn{1}{c}{\makecell{\textbf{InsightAgent}\\(GPT‑4o, $R=59$)}} \\
      \midrule
      Insufficient Quantitative Evidence 
          & \(12/15\) (80\%)  
          & \(10/15\) (67\%)  
          & \(8/30\) (27\%)  
          & \(13/59\) (22\%)  \\
      Limited Evidence Weighting 
        &  \(10/15\) (67\%)  
      &   \(8/15\) (53\%)  
      &  \(11/30\) (37\%)  
      &  \(18/59\) (31\%)  \\
      Omitted Heterogeneity Discussion 
      &  \(13/15\) (87\%)  
      &  \(10/15\) (67\%)  
      &  \(5/30\) (17\%)  
      &  \(7/59\) (11\%)  \\
      Hallucinations / Faithfulness Errors 
      &   \(8/15\) (53\%)  
      &   \(6/15\) (40\%)  
      &   \(6/30\) (20\%)  
      &   \(7/59\) (12\%)   \\
      \bottomrule
    \end{tabular}%
  }
\end{table*}

\section{Completed Systematic Reviews for Evaluation}
\label{appendix: sr-datasets}
We evaluated \sysname{} using 15 complete systematic reviews spanning diverse biomedical topics (Table~\ref{tab:studies}). All these reviews are publicly available and we replicate the corpus from PubMed, which are intended for research use. 

Each review’s search strategy was replicated according to recommended guidelines \cite{page2021prisma, chandler2019cochrane, medicine2011learning} to retrieve its original corpus of candidate articles, and the final set of included studies was recorded from the published review. The table summarizes each review’s publication year, title, total retrieved corpus size, and number of ultimately included articles. By systematically re-creating these 15 datasets, we ensure that our framework’s performance is assessed on authentic, rigorously vetted reviews, thereby enhancing the validity and comparability of our evaluation.

\begin{figure}[ht]
    \centering
    \includegraphics[width=0.8\linewidth]{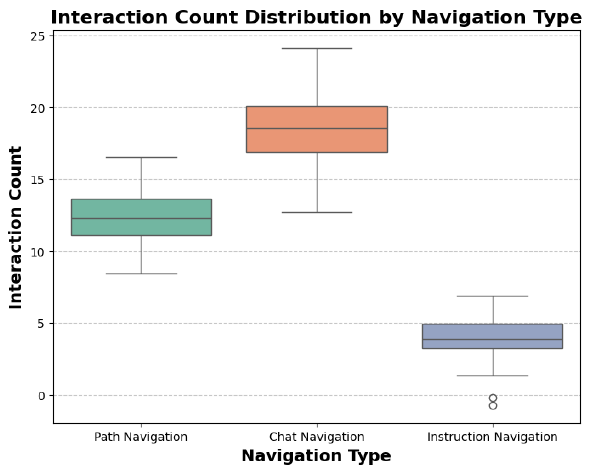}
    \caption{The number of interactions users conducted on average to complete systematic reviews across different navigation types.}
    \label{fig:interaction-dist}
\end{figure}

\section{Evaluation Rubric}
\label{appendix-evaluation-rubric}
Our evaluation rubric (Table~\ref{tab:rubric-cat1}, \ref{tab:rubric-cat2}, \ref{tab:rubric-cat3}, \ref{tab:rubric-cat4}, \ref{tab:rubric-cat5}) is developed by domain experts in biomedical research and guided by established systematic review frameworks \cite{page2021prisma}. To make is suitable for our evaluation setting, we tailor it to agent‐based summarization of a biomedical systematic review in extended‐abstract format. It evaluates five main dimensions: clarity of the research question and context, methods used (including data extraction and corpus coverage), completeness and accuracy of key biomedical findings, depth of discussion and practical takeaways, and overall writing quality. Each sub‐criterion outlines concise “full points” expectations and corresponding deductions, offering a detailed yet focused way to gauge how effectively an extended abstract captures core biomedical evidence and implications within the constraints of a shorter‐form document.

\section{Distribution of Interactions}
\label{appendix-interaction-stats}
We report the count of each type of interactions that users performed during the evaluation in Figure~\ref{fig:interaction-dist}

\section{Usability Questionnaire}
\label{appendix-usability-questionnair}
We conduct a user study inviting domain experts to perform a systematic review using \sysname{}, which integrates both a Radial Map and a Hierarchical Map for visualization and interaction. To assess the effectiveness of these designs, we formulate a set of tasks spanning cluster identification, path adjustment, and evidence synthesis navigation. Based on these tasks, and user's overall experience while using the system, we develop a structured questionnaire that covers five core usability categories: ease of use, confidence in recommendations, visualization-aided understanding, ability to guide or correct the agent, and overall satisfaction. \textbf{Table~\ref{tab:questionnairs}, ~\ref{tab:questionnairs-cont}} list all the questionnaire items, which allowed us to evaluate how effectively our visualizations and interactive features supported systematic review workflows.

\section{Reports Generated by human, \sysname{}, AutoSurvey and ChatCite }
\label{appendix-reports-with-evaluation}
We presents systematic review summaries generated by \sysname{}~(\textbf{Table~\ref{tab:report-insightagent}}), AutoSurvey~(\textbf{Table~\ref{tab:report-autosurvey}}) and ChatCite~(\textbf{Table~\ref{tab:report-chatcite}}), along with a human-conducted abstract reported in the orignal paper~\cite{tan2024among}~(\textbf{Table~\ref{tab:report-human}}). During the user study, participants leveraged 15 path navigation, 24 chat-based interactions, and 1 instruction navigation to refine the generated summary. 

\noindent \textbf{Comparison with Human-Curated Summary.}\\
\textbf{Table \ref{tab:report-insightagent}, \ref{tab:report-autosurvey}, \ref{tab:report-chatcite}} present the detailed evaluations based on our evaluation rubric (Appendix~\ref{appendix-evaluation-rubric}). The evaluation highlights that \sysname{}-produced reports captured essential findings with greater numerical specificity, while also offering structured synthesis across multiple perspectives.

Specifically, table~\ref{tab:report-insightagent} indicates that \sysname{} generated more extensive numeric details than the manually written review, an observation echoed by multiple participants. For instance, \sysname{} highlighted differential efficacy based on patient severity and time to treatment initiation, a nuanced perspective the human-curated report had touched upon only briefly. One expert remarked that \textit{“The system’s summary pinpoints how remdesivir’s impact may vary depending not just on disease severity but also treatment delay and patient demographics—a useful angle we hadn’t fully explored.”} \sysname{}’s structured approach also aligns closely with the human reference in terms of core conclusions, suggesting its capacity to synthesize abstract-level data effectively.

\textbf{AutoSurvey and ChatCite.}
Tables~\ref{tab:report-autosurvey}~and~\ref{tab:report-chatcite} show that while AutoSurvey retrieves multiple potentially relevant abstracts, it often introduces tangential discussions or unverified claims due to broad embedding-based retrieval. Clinical partners noted an “excess of irrelevant evidence” that diluted the final summary’s coherence. ChatCite, by contrast, provided shorter, more direct statements, yet lacked a cohesive structure suitable for systematic reviews, creating an “overwhelming” presentation of scattered facts. Participants found it challenging to assemble ChatCite’s unstructured bullet points into a narrative aligning with standard review guidelines.

\textbf{Strengths and Limitations.}
Overall, \sysname{} delivers a focused, multi‑perspective synthesis with greater numeric specificity—enabling rapid exploration of heterogeneous studies that traditionally demand extensive manual labor. In some cases, InsightAgent even surfaced more detailed effect estimates than the human‐written review, driven by domain experts interactively prompting for specific statistics to support their evolving queries. However, two key limitations remain. First, our reliance on abstract‐level inputs can omit critical numeric details—e.g., the WHO Solidarity trial’s mortality comparison (14.5\% vs.\ 15.6\%, $p=0.12$) and subgroup counts across nine RCTs totalling 13,085 patients were not fully captured (Table~\ref{tab:report-insightagent}-$E_1$, $E_2$). Second, without a dedicated statistical module, the system cannot compute confidence intervals or meta‑analytic effect sizes, limiting its ability to perform quantitative synthesis beyond reported values. Additional error categories, such as omitted heterogeneity discussion and hallucinations—are examined in detail in Appendix~\ref{appendix-detailed-error-analysis}. Future work should integrate structured data extraction and lightweight statistical analyses to enrich automatically generated summaries and further close the gap with human expert reviews.

\begin{figure*}[ht]
    \centering
    \includegraphics[width=\linewidth]{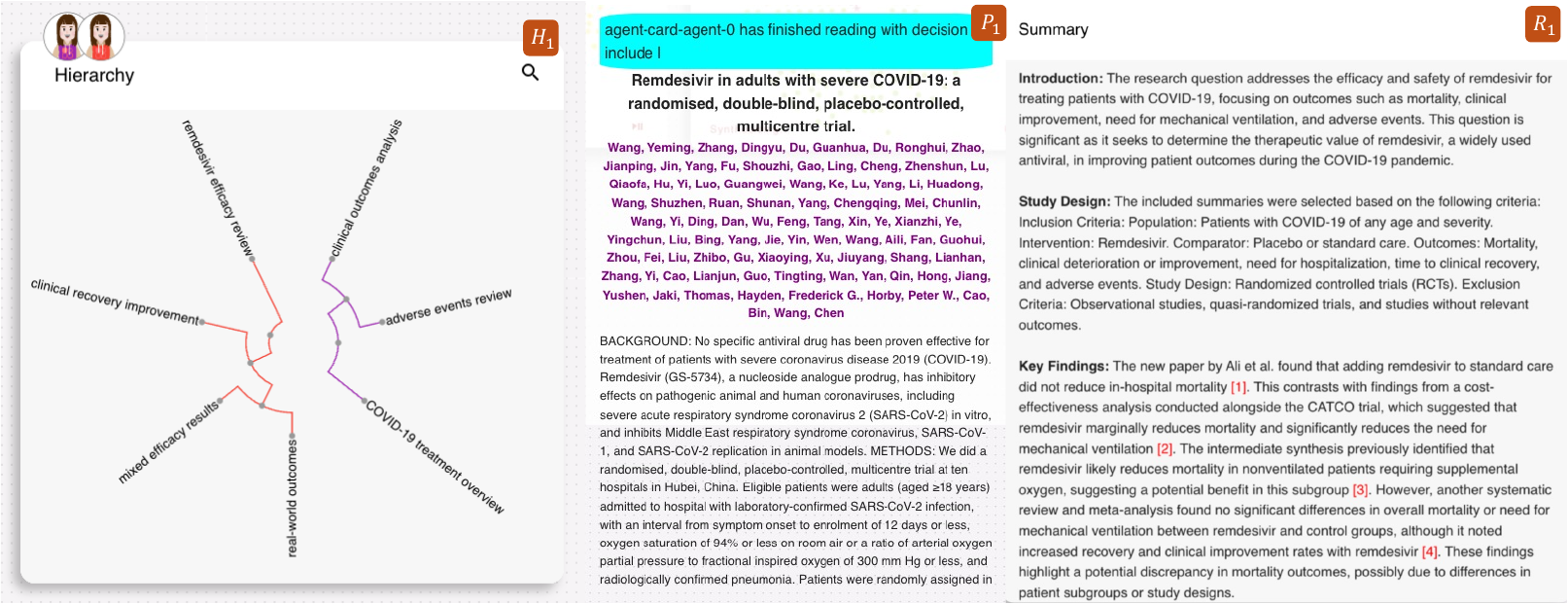}
    \caption{An integrated \emph{Synthesis View} in \sysname{} in an early stage of review, featuring two agents collaborating on the same systematic review. The \emph{Hierarchy} panel (left) merges each agent’s memory structure into a single visualization, color-coded to indicate which concepts originate from which agent. Hovering over a node reveals the underlying article or synthesized findings (e.g., \texttt{P1} for paper details, \texttt{R1} for an intermediate summary). This arrangement allows users to examine and reconcile multiple evidence streams in real time, comparing granular article content against higher-level synthesized conclusions.}
    \label{fig:insightagent}
\end{figure*}

\section{Detailed Error Analysis}
\label{appendix-detailed-error-analysis}
Below we define four principal error categories observed in system‐generated SR extended abstracts, describe how each is detected (via rubric subscores or expert feedback), and formalize its prevalence rate:
\[
  \mathrm{Rate}_e \;=\; \frac{N_e}{R}\,,
\]
where for each error type \(e\), \(N_e\) denotes the number of reports flagged under that category, and \(R\) denotes the total number of reports (see Table~\ref{tab:error-prevalence}).

\paragraph{1. Insufficient Statistical Analysis \& Quantitative Evidence} 
This error occurs when a summary omits concrete numerical synthesis (e.g., pooled effect sizes or incidence rates) that human reviewers would compute across studies. Detection is triggered if either the rubric subscore on item 3.3 (“Use of Quantitative or Specific Evidence”) is \(\leq 2\), or an expert explicitly notes missing statistics.  
As shown in Table~\ref{tab:error-prevalence}, this error affects 80\% of ChatCite and 67\% of AutoSurvey reports, but is reduced to 27\% for InsightAgent (Llama‑3.3) and 22\% for InsightAgent (GPT‑4o).  
\emph{Example feedback: “The summary notes improved symptom scores but fails to report the 25\% mean improvement and its confidence interval.”}

\paragraph{2. Limited Evidence Weighting} 
This error arises when the summary treats all studies equally, without accounting for design or quality differences (e.g., RCT vs.\ observational). We flag a report when rubric subscores on item 2.3 (“Coverage \& Representativeness”) or 2.5 (“Risk of Bias”) are \(\leq 2\), or an expert comment highlights uniform weighting.  
InsightAgent variants reduce this error from 67\% (ChatCite) and 53\% (AutoSurvey) to 37\% (Llama‑3.3) and 31\% (GPT‑4o).  
\emph{Example feedback: “All trials are summarized identically, but the large multi‐center RCT should carry greater emphasis than small cohort studies.”}

\paragraph{3. Omitted Heterogeneity Discussion} 
A report is flagged if it fails to acknowledge conflicting or subgroup findings (rubric item 3.4 “Variability / Heterogeneity” \(\leq 2\) or expert notes missing subgroup analysis). Baseline systems omit heterogeneity in 87\% (ChatCite) and 67\% (AutoSurvey) of cases, whereas InsightAgent lowers this to 17\% (Llama‑3.3) and 11\% (GPT‑4o).  
This reduction reflects how our interactive pipeline allows domain experts to prompt the agent for subgroup or variability checks during review, ensuring conflicting findings are surfaced before final synthesis.  
\emph{Example feedback: “There is no mention that remdesivir’s benefit differs by treatment delay or patient age, which emerged in several trials.”}

\paragraph{4. Hallucinations or Faithfulness Errors} 
This error covers any invented or misrepresented facts. We flag a report if rubric item 3.5 (“Faithfulness to Source Material”) is \(\leq 2\) or an expert identifies a discrepancy. Hallucinations occur in 53\% of ChatCite and 40\% of AutoSurvey summaries, but fall to 20\% (Llama‑3.3) and 12\% (GPT‑4o) with InsightAgent.  
The integrated evidence‐tracking graph and real‐time review controls enable users to verify each claim’s provenance on the fly, substantially limiting faithfulness errors.  
\emph{Example feedback: “The summary claims remdesivir reduced ICU stay by 50\%, yet no cited trial reports this figure.”}

By combining rubric‐based thresholds with expert annotations, we achieve both quantitative rigor and real‐world validity in pinpointing key failure modes of our agentic synthesis pipeline. The substantial reduction in omitted heterogeneity and faithfulness errors underscores the efficacy of human–in–the–loop oversight, while remaining error rates highlight avenues for enhancing statistical integration and bias mitigation.

\onecolumn
\FloatBarrier
\newcounter{boxcounter}
\begin{tcolorbox}[colback=gray!10, colframe=black!50, title=Retrieve Prompt]
\refstepcounter{boxcounter}  
\label{prompt-retrieves}
You are a biomedical research agent assisting with a systematic review on the question: \texttt{\{query\}}. 
You have access to a corpus of documents and must decide which papers to read \textit{during this iteration} based on their titles and initial abstracts. In doing so, you should consider whether each paper can contribute new evidence toward the research question.

Before this iteration, you have already read or processed the following papers:

\texttt{\{paper\_already\_read\}}

Below is a summary of any key findings you have discovered thus far:

\texttt{\{findings\_so\_far\}}

Here are the newly available papers for you to evaluate:

\texttt{\{available\_papers\}}

Please consult the following inclusion criteria to determine relevance:

\texttt{\{inclusion\_criteria\}}

Select one or more papers (by their indexes) if they provide additional, relevant information for the question \texttt{\{query\}} with detailed focus \texttt{\{detailed\_focus\}}. 

Otherwise, return \texttt{"skip"} if none appear pertinent or if they merely repeat information you have already processed.

Before making your selection, explain your reasoning and the rationale behind your choices.

Inspiration from prior conversation history is shown below:
$\{\text{inspiration\_conversation\_history}\}$

Your output must be in the JSON format below (and \textit{nothing else}):
\begin{verbatim}
```json
  {
    "thought": "<string>",
    "selected_papers": [
      // e.g., "1", "2", or "skip" if none are relevant
    ]
  }
```
\end{verbatim}
\end{tcolorbox}


\begin{tcolorbox}[colback=gray!10, colframe=black!50, title=Read Prompt]
\refstepcounter{boxcounter}  
\label{prompt-read}
You are a biomedical research agent assisting with a systematic review on the question: \texttt{\{query\}} with detailed focus \texttt{\{detailed\_focus\}}. 
Below, you will read a newly assigned paper, extracting any information that may relate to the review’s question. 

Before this iteration, you have processed:

\texttt{\{paper\_already\_read\}}

Key findings so far:

\texttt{\{findings\_so\_far\}}

The paper you need read is: 

\texttt{\{paper\_to\_read\}}

Be aware that this paper might not be relevant to \texttt{\{query\}} with the detailed focus \texttt{\{detailed\_focus\}}, or should be exclude considering the inclusion exclusion criteria: \texttt{\{inclusion\_criteria\}}. If it is relevant, produce an overall thought reflecting your updated understanding of the review, integrating what you have already discovered. 
If it is not relevant, explain briefly why you are excluding it.

If the content conflicts with previously read materials, do not resolve the conflict; simply include it in your overall thought for future exploration.

Consider the inspiration from prior conversation history with human before decision making:

$\{\text{inspiration\_conversation\_history}\}$

You must respond with a single JSON object, wrapped in an array, following the schema below. 

\begin{verbatim}
```json
  {
    "analysis": "<string>", 
    "response_preparation_analysis": "<string>",
    "related_to_query": true/false, 
    "reason_of_exclusion": "<string>",
    "summary_of_the_paper": "<string>",
    "summary_phrase": "<string>",
    "thought": "<string>"
  }
```
\end{verbatim}

Notes:

- analysis: includes all relevant details from the paper pertaining to the review, plus any interesting extra information. 

- response\_preparation\_analysis: how you intend to fulfill the user’s needs, given any prior instructions. 

- related\_to\_query: a boolean indicating if the paper addresses "{query}". 

- reason\_of\_exclusion: if \texttt{false} above, explain why. 

- summary\_of\_the\_paper: a concise overview of the content; return "not included" if no relevant details are found.

- summary\_phrase: a short, up-to-three-word phrase describing how the paper connects to "{query}". 

- thought: your overall updated understanding of the review, without extraneous details.

\end{tcolorbox}

\begin{tcolorbox}[colback=gray!10, colframe=black!50, title=Synthesize Prompt]
\refstepcounter{boxcounter}  
\label{prompt-synthesis}
You are a research agent focusing on a biomedical systematic review for the question \texttt{$\{\text{query}\}$}. You have just received a new paper summary and must integrate it with existing paper summaries or intermediate syntheses, if relevant. Your output should merge any overlapping or complementary information into a final synthesized summary as described below. 
Cite sources using \texttt{<citation>citation\_number</citation>} at the end of relevant sentences.

Here is the newly provided paper summary:

\texttt{$\{\text{current\_summary\_index}\}$}:\texttt{$\{\text{paper\_summary}\}$}

Below are previously generated summaries or intermediate syntheses, each with an ID:

\texttt{$\{\text{previous\_summaries}\}$}

You should:

[1] Identify the most relevant existing summary or synthesis to merge with this new paper. If none are relevant, clearly state why.  

[2] If a relevant item is another paper summary, combine them into a new intermediate synthesis. If it is an existing synthesis, add the new paper’s insights to that synthesis. If no match is found, explicitly explain that outcome.  

[3] Incorporate the user's summarization requirement \texttt{$\{\text{summarization\_requirement}\}$} 
into the final integrated summary if it applies to the content.

[4] Structure the final synthesized summary (plain text inside each section) with the following sections, and cite sources using \texttt{<citation>citation\_number</citation>}:

$======================$

• Introduction: Introduce the research question \texttt{$\{\text{query}\}$} and its broader context, explaining its significance.

• Study Design: Describe how the included summaries were chosen. Mention that the following inclusion-exclusion criteria were applied: \texttt{$\{\text{inclusion\_exclusion\_criteria}\}$}. Do not mention any search strategy.  

• Key Findings: Present core insights from all integrated evidence, highlighting any similarities, differences, patterns, or contradictions. Cite each source via \texttt{<citation>id</citation>} or \texttt{<citation>$\{\text{current\_summary\_index}\}$</citation>} for the new paper.  

• Conclusion: Summarize the overall outcomes, discussing whether they address the original question and any overarching patterns or implications.  

• Discussion: Reflect on the strength of the evidence, potential gaps, limitations, and suggest directions for future work. If certain aspects remain unclear, note that consulting the full text of some articles may be necessary.

$======================$

[5] Only use information provided in these summaries or syntheses; do not introduce details beyond the given text. Retain any citations from the identified synthesis if you are updating it.

After deciding which summaries or syntheses to include, return a single JSON object following the schema below, and nothing else:

\begin{verbatim}
{
  "identified_relevant_summaries": ["<id1>", "<id2>", ...],
  "reasoning": "<string>",
  "synthesized_summary": "<string>",
  "thought": "<string>"
}
\end{verbatim}

\noindent
Notes:

- \texttt{"identified\_relevant\_summaries"} is a list of IDs for any relevant summaries or empty if none match.

- \texttt{"reasoning"} explains why these items were selected (or not) in the context of \texttt{$\{\text{query}\}$}.

- \texttt{"synthesized\_summary"} is the new or updated synthesis, containing the five sections (Introduction, Study Design, Key Findings, Conclusion, Discussion) formatted in HTML with citations.

- \texttt{"thought"} is your overall perspective on \texttt{$\{\text{query}\}$} thus far, mindful of potential conflicts but without arbitrating them.

\end{tcolorbox}

\begin{tcolorbox}[colback=gray!10, colframe=black!50, title=Reflect Prompt]
\refstepcounter{boxcounter}  
\label{prompt-reflect}
You are a research agent assisting a human expert in conducting a systematic review for the question:
\texttt{\{\text{query}\}}. Below are the criteria you have been using to include or exclude studies:

\noindent
\texttt{\{\text{include\_exclude\_criteria}\}}

You have already read several papers and obtained certain findings (summaries or insights). The human expert
has now provided further input or questions. They may also have changed their guidance on which paper to read
(\texttt{\{\text{paper\_reading\_instruction\_if\_any}\}}), or revised the summarization requirement. If no
such specific instruction exists, this variable will be empty.

\noindent
\texttt{\{\text{findings\_so\_far}\}}\\
(Note: these findings represent your current overall insights.)

\noindent
\texttt{\{\text{conversation\_history}\}}\\
You must reflect on this new feedback or instruction to determine how your process should evolve. For instance:

- If the human suggests focusing on a different paper than you previously chose, you should note how your plan
  will adapt in the next iteration.

- If they introduce or change the summarization requirement, you should note the updates you will make.

- If they provide further critiques or clarifications, incorporate them into your plan.

- If they only express general approval or clarifications about your existing findings, you may continue without
  major changes.

\noindent
At the end of your reflection, you must produce a single JSON object, formatted following this schema:

\begin{verbatim}
```json
{
  "reflection": "<string>",
  "updates_on_additional_requirement": "<string>",
  "updates_on_criteria": "<string>",
  "updates_on_summarization_requirement": "<string>"
}
```
\end{verbatim}
Note: 

-\texttt{``reflection"} is your overall reasoning on how to modify your process or maintain it, based on the new human feedback.

-\texttt{``updates\_on\_additional\_requirement"} describes any further research directions or sub-questions you plan to pursue at the user’s request; if there is no update, leave this empty.

-\texttt{``updates\_on\_criteria"} indicates any changes to your inclusion/exclusion criteria after reflecting on human critiques; if none, leave this empty.

-\texttt{``updates\_on\_summarization\_requirement"} details any shifts in how you will summarize information in subsequent steps (e.g., more concise, focusing on specific methodology or outcome), or remains empty if no change is needed.

\end{tcolorbox}

\FloatBarrier

\FloatBarrier

\begin{table*}
\centering
\small 
\begin{tabularx}{\textwidth}{lXl} 
\toprule
\textbf{Year} & \textbf{Title} & \textbf{Corpus/Inclusion} \\
\midrule
2024 & A systematic review on the impact of sub-epidermal \cite{avsar2024systematic} &
  712 Total / 10 Included \\
2024 & Artificial Intelligence Support for Informal Patient Caregivers \cite{borna2024artificial} &
  381 Total / 10 Included \\
2024 & Nutritional interventions for preventing \cite{langer2024nutritional} &
  72 Total / 23 Included \\
2024 & Community dwelling older people \cite{iida2025community} &
  455 Total / 5 Included \\
2024 & Prevalence of persistent hypertension following pregnancy \cite{mukosha2024prevalence} &
  752 Total / 22 Included \\
2024 & Insulin Delivery Technology for Treatment in Infants \cite{panza2024insulin}&
  204 Total / 22 Included \\
2024 & A Systematic Review of Nurses' Perceptions \cite{park2024systematic} &
  449 Total / 21 Included \\
2024 & Text Messaging for Medication Adherence in CVD Prevention \cite{hyun2024mobile} &
  4,162 Total / 18 Included \\
2024 & Psychological interventions for asthma in children and adolescents \cite{sharrad2024psychological} &
  6,665 Total / 34 Included \\
2024 & Cognitive mental workload of emergency nursing \cite{surendran2024cognitive} &
  353 Total / 57 Included \\
2024 & Remdesivir for Patients with COVID-19 \cite{tan2024among} &
  491 Total / 9 Included \\
2024 & The prevalence of pulmonary hypertension in post-tuberculosis \cite{van2024prevalence} &
  542 Total / 34 Included \\
2022 & Effects of Caffeine Intake on endurance \cite{wang2022effects} &
  1,107 Total / 21 Included \\
2024 & Evening versus morning dosing regimen \cite{wu2024evening} &
  7,356 Total / 27 Included \\
2023 & Efficacy of single-pill combination \cite{xie2023efficacy} &
  1,697 Total / 32 Included \\
\bottomrule
\end{tabularx}
\caption{ 15 completed systematical reviews used in evaluation, including publication year, review title, total retrieved corpus, and final number of included articles.}
\label{tab:studies}
\end{table*}

\FloatBarrier

\begin{table*}[t]
\centering
\small
\begin{tabularx}{\textwidth}{X}
\toprule
\textbf{Detailed Evaluation Rubric -- Category 1: Introduction \& Research Context (15 points)} \\
\midrule
These items evaluate how well the extended abstract states the research question, provides justification, and outlines clear objectives for practice or research. \\[6pt]

\begin{itemize}[topsep=0pt, partopsep=0pt, parsep=0pt, itemsep=0pt]
  \item 1.1 Clarity of Research Question (5 points)
  \begin{itemize}[topsep=0pt, itemsep=0pt]
    \item \textit{Full Points (5):}
    \begin{itemize}[topsep=0pt, itemsep=0pt]
      \item The extended abstract explicitly states the research question (e.g., comparing interventions X and Y).
      \item The scope is clear (population, condition, outcomes).
    \end{itemize}
    \item \textit{Deductions:}
    \begin{itemize}[topsep=0pt, itemsep=0pt]
      \item (-1 \textasciitilde 2 points) if the question is vaguely stated or missing key details.
      \item (-3 \textasciitilde 4 points) if the research question is unclear or scattered.
      \item (-5 points) if the question is not stated at all or is entirely irrelevant.
    \end{itemize}
  \end{itemize}

  \item 1.2 Justification \& Relevance (5 points)
  \begin{itemize}[topsep=0pt, itemsep=0pt]
    \item \textit{Full Points (5):}
    \begin{itemize}[topsep=0pt, itemsep=0pt]
      \item Provides a compelling reason for why this extended abstract is necessary (e.g., gaps in existing knowledge).
      \item Shows clear alignment with user‐provided context or corpus.
    \end{itemize}
    \item \textit{Deductions:}
    \begin{itemize}[topsep=0pt, itemsep=0pt]
      \item (-1 \textasciitilde 2 points) if the justification is too brief or lacks evidence.
      \item (-3 \textasciitilde 4 points) if little rationale is given and significance is unclear.
      \item (-5 points) if no justification is provided at all.
    \end{itemize}
  \end{itemize}

  \item 1.3 Objectives \& Significance (5 points)
  \begin{itemize}[topsep=0pt, itemsep=0pt]
    \item \textit{Full Points (5):}
    \begin{itemize}[topsep=0pt, itemsep=0pt]
      \item Clearly identifies intended impact (practice, policy, or future research).
      \item States how the agent‐based summary could help clinicians, researchers, or stakeholders.
    \end{itemize}
    \item \textit{Deductions:}
    \begin{itemize}[topsep=0pt, itemsep=0pt]
      \item (-1 \textasciitilde 2 points) if objectives are too general or only partially linked.
      \item (-3 \textasciitilde 4 points) if objectives are not mentioned or lack practical implications.
      \item (-5 points) if it is entirely missing or irrelevant.
    \end{itemize}
  \end{itemize}
\end{itemize} \\[3pt]

\bottomrule
\end{tabularx}
\caption{Detailed Evaluation Rubric -- Category 1}
\label{tab:rubric-cat1}
\end{table*}

\begin{table*}[t]
\centering
\small
\begin{tabularx}{\textwidth}{X}
\toprule
\textbf{Detailed Evaluation Rubric -- Category 2: Summarization Approach \& Method (25 points)} \\
\midrule
These items focus on how the agent or system generated the extended abstract, inclusion/exclusion criteria, data extraction, and potential biases. Also adapted from traditional methodology to reflect the user-provided corpus. \\[6pt]

\begin{itemize}[topsep=0pt, partopsep=0pt, parsep=0pt, itemsep=0pt]
  \item 2.1 Agent-Based Summarization Description (5 points)
  \begin{itemize}[topsep=0pt, itemsep=0pt]
    \item \textit{Full Points (5):}
    \begin{itemize}[topsep=0pt, itemsep=0pt]
      \item Describes how the system generated the abstract (e.g., from user corpus, rules, filters).
      \item Mentions the basic workflow (e.g., “The agent filtered 50 articles…”).
    \end{itemize}
    \item \textit{Deductions:}
    \begin{itemize}[topsep=0pt, itemsep=0pt]
      \item (-1 \textasciitilde 2 points) if the system’s role is only vaguely mentioned.
      \item (-3 \textasciitilde 4 points) if the method is unclear or incomplete.
      \item (-5 points) if no method is described at all.
    \end{itemize}
  \end{itemize}

  \item 2.2 Inclusion/Exclusion Rationale \& Data Sources (5 points)
  \begin{itemize}[topsep=0pt, itemsep=0pt]
    \item \textit{Full Points (5):}
    \begin{itemize}[topsep=0pt, itemsep=0pt]
      \item Briefly states what types of articles or data were included (e.g., RCTs, observational) and any user-driven filters.
      \item Mentions the source of the corpus (e.g., “User uploaded 50 full-text articles from PubMed…”).
    \end{itemize}
    \item \textit{Deductions:}
    \begin{itemize}[topsep=0pt, itemsep=0pt]
      \item (-1 \textasciitilde 2 points) if criteria or corpus sources are vague or incomplete.
      \item (-3 \textasciitilde 4 points) if it’s unclear what data was included/excluded.
      \item (-5 points) if no mention of data sources or selection criteria is made.
    \end{itemize}
  \end{itemize}

  \item 2.3 Coverage \& Representativeness (5 points)
  \begin{itemize}[topsep=0pt, itemsep=0pt]
    \item \textit{Full Points (5):}
    \begin{itemize}[topsep=0pt, itemsep=0pt]
      \item Demonstrates that the abstract captures the key range of the user‐provided corpus.
      \item Incorporates relevant findings from diverse studies (if applicable).
    \end{itemize}
    \item \textit{Deductions:}
    \begin{itemize}[topsep=0pt, itemsep=0pt]
      \item (-1 \textasciitilde 2 points) if only a narrow subset is summarized without justification.
      \item (-3 \textasciitilde 4 points) if major, important studies/themes are missing.
      \item (-5 points) if it’s unclear the summary reflects the corpus at all.
    \end{itemize}
  \end{itemize}

  \item 2.4 Data Extraction \& Reliability (5 points)
  \begin{itemize}[topsep=0pt, itemsep=0pt]
    \item \textit{Full Points (5):}
    \begin{itemize}[topsep=0pt, itemsep=0pt]
      \item Explains how key info (outcomes, participants, etc.) was extracted or verified.
      \item Mentions how errors/biases in extraction might be handled.
    \end{itemize}
    \item \textit{Deductions:}
    \begin{itemize}[topsep=0pt, itemsep=0pt]
      \item (-1 \textasciitilde 2 points) if extraction process is vaguely described.
      \item (-3 \textasciitilde 4 points) if there is no clarity on relevant details.
      \item (-5 points) if steps are absent or contradictory.
    \end{itemize}
  \end{itemize}

  \item 2.5 Risk of Bias / Limitations (5 points)
  \begin{itemize}[topsep=0pt, itemsep=0pt]
    \item \textit{Full Points (5):}
    \begin{itemize}[topsep=0pt, itemsep=0pt]
      \item Acknowledges potential weaknesses (e.g., model may overlook nuances).
      \item Identifies any known biases in the agent’s approach or corpus.
    \end{itemize}
    \item \textit{Deductions:}
    \begin{itemize}[topsep=0pt, itemsep=0pt]
      \item (-1 \textasciitilde 2 points) if risk of bias is only superficially noted.
      \item (-3 \textasciitilde 4 points) if major limitations are ignored.
      \item (-5 points) if no mention of limitations or bias at all.
    \end{itemize}
  \end{itemize}
\end{itemize} \\[3pt]

\bottomrule
\end{tabularx}
\caption{Detailed Evaluation Rubric -- Category 2}
\label{tab:rubric-cat2}
\end{table*}

\begin{table*}[t]
\centering
\small
\begin{tabularx}{\textwidth}{X}
\toprule
\textbf{Detailed Evaluation Rubric -- Category 3: Comprehensiveness \& Accuracy of Summaries (30 points)} \\
\midrule
These items focus on how thoroughly and accurately the extended abstract covers all major findings, uses quantitative evidence, and addresses heterogeneity or conflicting data. \\[6pt]

\begin{itemize}[topsep=0pt, partopsep=0pt, parsep=0pt, itemsep=0pt]
  \item 3.1 Relevance \& Completeness (10 points)
  \begin{itemize}[topsep=0pt, itemsep=0pt]
    \item \textit{Full Points (10):}
    \begin{itemize}[topsep=0pt, itemsep=0pt]
      \item Accurately reflects all major findings relevant to the research question.
      \item Notes key outcomes, interventions, or theories from the corpus.
    \end{itemize}
    \item \textit{Deductions:}
    \begin{itemize}[topsep=0pt, itemsep=0pt]
      \item (-1 \textasciitilde 3 points) if some significant findings are missing or incomplete.
      \item (-4 \textasciitilde 7 points) if large sections are omitted, misleading the summary.
      \item (-8 \textasciitilde 10 points) if it fails to capture the essence entirely.
    \end{itemize}
  \end{itemize}

  \item 3.2 Clarity \& Cohesion of Findings (5 points)
  \begin{itemize}[topsep=0pt, itemsep=0pt]
    \item \textit{Full Points (5):}
    \begin{itemize}[topsep=0pt, itemsep=0pt]
      \item Information flows logically, with clear transitions; no abrupt contradictions.
    \end{itemize}
    \item \textit{Deductions:}
    \begin{itemize}[topsep=0pt, itemsep=0pt]
      \item (-1 \textasciitilde 2 points) if transitions are unclear or haphazard.
      \item (-3 \textasciitilde 5 points) if the text is difficult to follow or contradictory.
    \end{itemize}
  \end{itemize}

  \item 3.3 Use of Quantitative or Specific Evidence (5 points)
  \begin{itemize}[topsep=0pt, itemsep=0pt]
    \item \textit{Full Points (5):}
    \begin{itemize}[topsep=0pt, itemsep=0pt]
      \item Includes specific data points or effect sizes where possible.
      \item Avoids vague statements like “Most studies are positive” without data.
    \end{itemize}
    \item \textit{Deductions:}
    \begin{itemize}[topsep=0pt, itemsep=0pt]
      \item (-1 \textasciitilde 2 points) if only partial specificity or missing effect sizes.
      \item (-3 \textasciitilde 4 points) if mostly qualitative, omitting key quantitative data.
      \item (-5 points) if no concrete data despite availability.
    \end{itemize}
  \end{itemize}

  \item 3.4 Discussion of Variability / Heterogeneity (5 points)
  \begin{itemize}[topsep=0pt, itemsep=0pt]
    \item \textit{Full Points (5):}
    \begin{itemize}[topsep=0pt, itemsep=0pt]
      \item Addresses differences among studies and acknowledges inconsistent findings.
    \end{itemize}
    \item \textit{Deductions:}
    \begin{itemize}[topsep=0pt, itemsep=0pt]
      \item (-1 \textasciitilde 2 points) if variability is briefly mentioned but not explained.
      \item (-3 \textasciitilde 4 points) if contradictory findings lack context.
      \item (-5 points) if no mention of heterogeneity at all.
    \end{itemize}
  \end{itemize}

  \item 3.5 Faithfulness to Source Material (5 points)
  \begin{itemize}[topsep=0pt, itemsep=0pt]
    \item \textit{Full Points (5):}
    \begin{itemize}[topsep=0pt, itemsep=0pt]
      \item Conclusions and claims clearly stem from the user‐provided corpus.
      \item No major distortions or misinterpretations of original study conclusions.
    \end{itemize}
    \item \textit{Deductions:}
    \begin{itemize}[topsep=0pt, itemsep=0pt]
      \item (-1 \textasciitilde 2 points) if minor discrepancies or paraphrasing errors occur.
      \item (-3 \textasciitilde 4 points) if some interpretations conflict with source data.
      \item (-5 points) if the summary significantly misrepresents the corpus or invents data.
    \end{itemize}
  \end{itemize}
\end{itemize} \\[3pt]

\bottomrule
\end{tabularx}
\caption{Detailed Evaluation Rubric -- Category 3}
\label{tab:rubric-cat3}
\end{table*}

\begin{table*}[t]
\centering
\small
\begin{tabularx}{\textwidth}{X}
\toprule
\textbf{Detailed Evaluation Rubric -- Category 4: Discussion, Implications \& Future Directions (20 points)} \\
\midrule
These items evaluate how well the summary interprets findings, gives practical recommendations, and proposes future research. \\[6pt]

\begin{itemize}[topsep=0pt, partopsep=0pt, parsep=0pt, itemsep=0pt]
  \item 4.1 Interpretation of Findings (10 points)
  \begin{itemize}[topsep=0pt, itemsep=0pt]
    \item \textit{Full Points (10):}
    \begin{itemize}[topsep=0pt, itemsep=0pt]
      \item Interprets the summarized data in context (e.g., “These findings suggest X is more effective…”).
      \item Identifies overarching trends, limitations, and potential biases clearly.
    \end{itemize}
    \item \textit{Deductions:}
    \begin{itemize}[topsep=0pt, itemsep=0pt]
      \item (-1 \textasciitilde 3 points) if interpretation is present but lacks clarity or depth.
      \item (-4 \textasciitilde 7 points) if major outcomes are not linked to broader discussion.
      \item (-8 \textasciitilde 10 points) if there is little or no interpretation, or contradictions.
    \end{itemize}
  \end{itemize}

  \item 4.2 Practical Implications or Recommendations (5 points)
  \begin{itemize}[topsep=0pt, itemsep=0pt]
    \item \textit{Full Points (5):}
    \begin{itemize}[topsep=0pt, itemsep=0pt]
      \item Offers thoughtful suggestions for clinicians, policymakers, or researchers.
      \item Ties the summarized evidence back to real‐world application.
    \end{itemize}
    \item \textit{Deductions:}
    \begin{itemize}[topsep=0pt, itemsep=0pt]
      \item (-1 \textasciitilde 2 points) if implications are only briefly stated.
      \item (-3 \textasciitilde 4 points) if recommendations are vague or lack linkage to data.
      \item (-5 points) if no discussion of implications is provided.
    \end{itemize}
  \end{itemize}

  \item 4.3 Suggestions for Future Research (5 points)
  \begin{itemize}[topsep=0pt, itemsep=0pt]
    \item \textit{Full Points (5):}
    \begin{itemize}[topsep=0pt, itemsep=0pt]
      \item Identifies gaps or unresolved questions (e.g., limited data on long‐term effects of X).
      \item Proposes clear avenues for deeper investigation.
    \end{itemize}
    \item \textit{Deductions:}
    \begin{itemize}[topsep=0pt, itemsep=0pt]
      \item (-1 \textasciitilde 2 points) if future directions are superficial.
      \item (-3 \textasciitilde 4 points) if they are barely mentioned or unconnected to corpus.
      \item (-5 points) if no future research direction is provided at all.
    \end{itemize}
  \end{itemize}
\end{itemize} \\[3pt]

\bottomrule
\end{tabularx}
\caption{Detailed Evaluation Rubric -- Category 4}
\label{tab:rubric-cat4}
\end{table*}

\begin{table*}[t]
\centering
\small
\begin{tabularx}{\textwidth}{X}
\toprule
\textbf{Detailed Evaluation Rubric -- Category 5: Writing Quality, Fluency \& Referencing (10 points)} \\
\midrule
These items assess clarity, coherence, and proper referencing or attribution in the extended abstract. \\[6pt]

\begin{itemize}[topsep=0pt, partopsep=0pt, parsep=0pt, itemsep=0pt]
  \item 5.1 Readability \& Structure (5 points)
  \begin{itemize}[topsep=0pt, itemsep=0pt]
    \item \textit{Full Points (5):}
    \begin{itemize}[topsep=0pt, itemsep=0pt]
      \item The text is well‐organized, concise, and easy to understand.
      \item Logical headings or paragraph breaks guide the reader.
    \end{itemize}
    \item \textit{Deductions:}
    \begin{itemize}[topsep=0pt, itemsep=0pt]
      \item (-1 \textasciitilde 2 points) if minor clarity issues occur sporadically.
      \item (-3 \textasciitilde 4 points) if the writing is frequently unclear or structurally confusing.
      \item (-5 points) if the writing is so poor that the extended abstract is barely comprehensible.
    \end{itemize}
  \end{itemize}

  \item 5.2 Referencing \& Supporting Evidence (5 points)
  \begin{itemize}[topsep=0pt, itemsep=0pt]
    \item \textit{Full Points (5):}
    \begin{itemize}[topsep=0pt, itemsep=0pt]
      \item References (or placeholders) are used where the agent draws on specific studies or data.
      \item Citations are relevant; any direct quotations are properly attributed.
    \end{itemize}
    \item \textit{Deductions:}
    \begin{itemize}[topsep=0pt, itemsep=0pt]
      \item (-1 \textasciitilde 2 points) if references are incomplete or inconsistent.
      \item (-3 \textasciitilde 4 points) if missing references for major claims.
      \item (-5 points) if there are no references at all in contexts that need them.
    \end{itemize}
  \end{itemize}
\end{itemize} \\[3pt]

\bottomrule
\end{tabularx}
\caption{Detailed Evaluation Rubric -- Category 5}
\label{tab:rubric-cat5}
\end{table*}
\FloatBarrier

\begin{table*}[t]
\centering
\small
\begin{tabularx}{\textwidth}{X}
\toprule
\textbf{Detailed Questionnaires after User Study } \\
\midrule
\textbf{Category 1: The System Was Easy to Use} \\
These questions evaluate how easily participants can navigate 
and operate different components of \sysname{}. \\
\begin{enumerate}[topsep=0pt, partopsep=0pt, parsep=0pt, itemsep=0pt]
  \item Radial Chart – Search \& Navigation
  \begin{itemize}[topsep=0pt, partopsep=0pt, parsep=0pt, itemsep=0pt]
      \item How easy was it to locate the target article using the search function?
      
      Scale: (1 = Very difficult, 5 = Very easy)
  \end{itemize}
  \item Hierarchical Map – Exploration
  \begin{itemize}
      \item How easy was it to explore and understand the synthesis process?

      Scale: (1 = Very difficult, 5 = Very easy)
  \end{itemize}

  \item Hierarchical Map – Tracing Citations
     \begin{itemize}
         \item How easy was it to trace the citation path for the summary?
            
        Scale: (1 = Very difficult, 5 = Very easy)
   
     \end{itemize}

    \item Hierarchical Map – Locating Specific Evidence
    \begin{itemize}
        \item How easy was it to locate specific evidence within the Hierarchical Map?

Scale: (1 = Very difficult, 5 = Very easy)
    \end{itemize}
\end{enumerate}
\\\hline
\textbf{Category 2: Confidence in Recommendations} \\
These questions measure the participant’s trust in the system’s outputs, their own confidence in interpreting the agent’s actions, and overall trust in the final systematic review.
 \\
\begin{enumerate}[topsep=0pt, partopsep=0pt, parsep=0pt, itemsep=0pt]
  \item Radial Chart – Agent’s Progress
  \begin{itemize}[topsep=0pt, partopsep=0pt, parsep=0pt, itemsep=0pt]
      \item How confident are you in your understanding of the agent’s progress?
      
      Scale: (1 = Not confident at all, 5 = Very confident)
  \end{itemize}
  \item Hierarchical Map – Source Identification
  \begin{itemize}
      \item How confident are you that you identified all the original sources correctly?

      Scale: (1 = Not confident at all, 5 = Very confident)
  \end{itemize}

  \item Hierarchical Map – Agent Incorporating Feedback
     \begin{itemize}
         \item How confident are you that the agent will incorporate your feedback into future summaries?
            
        Scale: (1 = Not confident at all, 5 = Very confident)
   
     \end{itemize}

    \item Hierarchical Map – Synthesis Accuracy
    \begin{itemize}
        \item How confident are you that the agent is synthesizing the correct information?

Scale: Scale: (1 = Not confident at all, 5 = Very confident)
    \end{itemize}

\end{enumerate}

\\\hline
\textbf{Category 3: Visualization-Aided Understanding} \\
These items assess how well the visual interfaces (Radial Chart and Hierarchical Map) help users understand cluster boundaries, topic coverage, citation structures, and the synthesis process.
 \\
\begin{enumerate}[topsep=0pt, partopsep=0pt, parsep=0pt, itemsep=0pt]
  \item Radial Chart – Cluster Clarity
  \begin{itemize}[topsep=0pt, partopsep=0pt, parsep=0pt, itemsep=0pt]
      \item How clear were the cluster boundaries on the Radial Chart?
      
      Scale: (1 = Not clear at all, 5 = Very clear)
  \end{itemize}
  \item Radial Chart – Topic Identification
  \begin{itemize}
      \item How easy was it to understand the main topics of each cluster?

      Scale: (1 = Very difficult, 5 = Very easy)
  \end{itemize}

  \item Radial Chart – Node Information
     \begin{itemize}
         \item How clear was the information provided when hovering over the article node?
            
        Scale: (1 = Not clear at all, 5 = Very clear)
   
     \end{itemize}

    \item Radial Chart – Topic Coverage
    \begin{itemize}
        \item How comprehensive was the Radial Chart in displaying topic coverage?

Scale: (1 = Not comprehensive at all, 5 = Very comprehensive)
    \end{itemize}

    \item Radial Chart – Missing Areas
    \begin{itemize}
        \item How easy was it to identify missing areas in the agent’s coverage?

Scale: (1 = Very difficult, 5 = Very easy)
    \end{itemize}

    \item Hierarchical Map – Synthesis Structure
    \begin{itemize}
        \item How clear was the hierarchical structure of the evidence synthesis?

Scale: (1 = Not clear at all, 5 = Very clear)
    \end{itemize}
    \item Hierarchical Map – Agent’s Synthesis Process
    \begin{itemize}
        \item How clear was the agent’s synthesis process as displayed in the Hierarchical Map?

Scale: (1 = Not clear at all, 5 = Very clear)
    \end{itemize}


\end{enumerate}
\\\hline  
\bottomrule
\end{tabularx}
\caption{Detailed Questionnaire}
\label{tab:questionnairs}
\end{table*}

\begin{table*}[t]
\centering
\small
\begin{tabularx}{\textwidth}{X}
\toprule
\textbf{Detailed Questionnaires after User Study } \\
\midrule
\textbf{Category 4: Ability to Guide or Correct Agents} \\
These questions measure how effectively participants can detect and fix issues, give feedback, or redirect the agent using the system’s interfaces. \\
\begin{enumerate}[topsep=0pt, partopsep=0pt, parsep=0pt, itemsep=0pt]
  \item Radial Chart – Path Correction
  \begin{itemize}[topsep=0pt, partopsep=0pt, parsep=0pt, itemsep=0pt]
      \item How easy was it to identify incorrect paths on the Radial Chart?
      
      Scale: (1 = Very difficult, 5 = Very easy)
  \end{itemize}
  \item Radial Chart – Drag-and-Drop
  \begin{itemize}
      \item How intuitive was the drag-and-drop functionality for adjusting paths?

      Scale: (1 = Not intuitive at all, 5 = Very intuitive)
  \end{itemize}

  \item Hierarchical Map – Missing Citations
     \begin{itemize}
         \item How easy was it to identify missing citations or incorrect paths?
            
        Scale: (1 = Very difficult, 5 = Very easy)
     \end{itemize}

    \item Hierarchical Map – Feedback on Summaries
    \begin{itemize}
        \item How intuitive was the feedback process within the Hierarchical Map?

Scale: (1 = Not intuitive at all, 5 = Very intuitive)
    \end{itemize}
\end{enumerate}
\\\hline
\textbf{Category 5: Overall Satisfaction} \\
These questions capture general satisfaction and user impressions of how well the system performs and meets user needs.
 \\
\begin{enumerate}[topsep=0pt, partopsep=0pt, parsep=0pt, itemsep=0pt]
  
  \item Overall System Trustworthiness
    \begin{itemize}
        \item To what extent do you trust the generated systematic review?

Scale: (1 = Not confident at all, 5 = Very confident)
    \end{itemize}
\end{enumerate}
\\\hline  
\bottomrule
\end{tabularx}
\caption{Detailed Questionnaire CONT.}
\label{tab:questionnairs-cont}
\end{table*} 
\begin{table*}[t]
\centering
\small
\caption{Report generated by \textbf{Human (original report)} on Systematic review for Remdesivir be Used for Treatment \cite{tan2024among} [\preferred{preferred statements are highlighted in green}, \lesspreferred{less preferred statements are highlighted in orange}, \notpreferred{not preferred statements are highlighted in red}]}
\label{tab:report-human}
\begin{tabularx}{\textwidth}{X}
\toprule
\textbf{Human} \\
\midrule
\textbf{Background}. Remdesivir is an intravenously administered antiviral drug that inhibits RNA-dependent RNA polymerase. In vitro studies have shown that remdesivir can inhibit the growth of the COVID-19 virus in infected Vero cells and can inhibit infection in human cell lines.

\textbf{Objective}. To determine the efficacy and safety of remdesivir in treating patients with COVID-19 infection.

\textbf{Methods}. A systematic search of electronic medical literature databases was done from inception until September 4, 2022. Search for ongoing studies and preprints was also done. Risk of bias assessment was done using Cochrane risk of bias tool version 2.0. Measures of effect used were relative risk (RR) and 95\% confidence interval (CI). Subgroup analysis by disease severity was preplanned. The estimates for efficacy and safety of remdesivir was calculated using Review Manager 5.4 software.

\textbf{Results}. \preferred{Nine randomized controlled trials with 13,085 participants were identified(\textbf{Report a detailed number of participants / patients in these studies})}. Eight of the included studies recruited confirmed COVID-19 patients needing hospitalization, while one study limited recruitment to nonhospitalized patients. \preferred{Remdesivir showed significant benefit for outpatients with mild to moderate disease with at least one risk factor for disease progression in terms of COVID 19-related hospitalization (RR 0.13 95\% CI 0.03 to 0.59), all-cause hospitalization (RR 0.28, 95\% CI 0.10 to 0.75), and need for medically-attended visits (RR 0.19, 95\% CI 0.07 to 0.56).(\textbf{detailed synthesized statistics})} \\ 
\textbf{Discussion}. Remdesivir demonstrates benefit in patients with mild, moderate, and severe COVID-19 but not in critical cases requiring mechanical ventilation. Its safety profile aligns with international recommendations. \\ 
\textbf{Conclusion}. Current evidence supports the use of remdesivir in selected COVID-19 patients.\\
\bottomrule
\end{tabularx}
\end{table*} 


\begin{table*}[t]
\centering
\small
\caption{Report generated by \textbf{\sysname{}} on Systematic review for Remdesivir be Used for Treatment \cite{tan2024among} [\preferred{preferred statements are highlighted in green}, \lesspreferred{less preferred statements are highlighted in orange}, \notpreferred{not preferred statements are highlighted in red}]} 
\label{tab:report-insightagent}
\begin{tabularx}{\textwidth}{X}
\toprule
\textbf{\sysname{}(GPT-4o)} \\
\midrule
\textbf{Introduction:}
Remdesivir has been widely used as an antiviral treatment for COVID-19, but its effectiveness remains debated. This review evaluates the efficacy and safety of remdesivir in treating COVID-19 patients, focusing on outcomes such as mortality, clinical improvement, mechanical ventilation, and adverse events. The review synthesizes findings from multiple randomized controlled trials (RCTs) to provide a comprehensive assessment of its clinical impact.

\textbf{Study Design:}
\lesspreferred{$^{E_2}$This systematic review analyzed 491 studies, of which 11 met the inclusion criteria for final synthesis. These studies were RCTs comparing remdesivir with placebo or standard care in patients with COVID-19 of varying severity (\textbf{More detailed number of participants in these studies would make it more convincing})}. The inclusion criteria required studies to assess mortality, clinical improvement, hospitalization, time to recovery, or adverse events. Observational studies, quasi-randomized trials, and those without relevant outcomes were excluded. The included trials span different populations and care settings, allowing for a broad assessment of remdesivir’s effectiveness [1-11].

\textbf{Key Findings:}

\preferred{\textbf{Mortality Impact: (preferred discussion topic for findings)}}
The WHO Solidarity trial found no significant reduction in mortality among hospitalized COVID-19 patients treated with \lesspreferred{$^{E_1}$ remdesivir (14.5\% vs. 15.6\%, RR 0.91, p=0.12) [2]. A meta-analysis combining multiple trials showed a modest mortality reduction in non-ventilated patients (RR 0.86, p=0.02) (\textbf{More synthesized stats result is preferred})} but no benefit for ventilated patients [2]. The NOR-Solidarity trial reported similar mortality rates between remdesivir and standard care groups [4].

\preferred{\textbf{Clinical Improvement \& Recovery Time: (preferred discussion topic for findings)}}
The DisCoVeRy trial found no substantial improvement in clinical outcomes with remdesivir treatment [1]. The PINETREE study reported that early remdesivir use in high-risk outpatients significantly reduced hospitalization or death (HR 0.13, p=0.008) [3]. The Spinner et al. trial observed that a 5-day remdesivir course resulted in a statistically significant improvement in clinical status, whereas the 10-day regimen showed no clear benefit [6]. 

\textbf{Need for Mechanical Ventilation:}
The WHO Solidarity trial found that remdesivir reduced the progression to mechanical ventilation in non-ventilated patients (RR 0.88, p=0.04) [2]. In ventilated patients, remdesivir did not improve survival or clinical outcomes (RR 1.13, p=0.32) [2].

\preferred{\textbf{Adverse Events \& Safety Profile: (preferred discussion topic for findings)}}
The NOR-Solidarity trial found no significant difference in adverse events between remdesivir and control groups [4]. Common adverse effects associated with remdesivir included nausea, hypokalemia, and headache, though these were generally mild [6]. Remdesivir was linked to a slight delay in hospital discharge due to its intravenous administration requirement [2].

\textbf{Conclusion:}
\preferred{Remdesivir provides modest clinical benefits in non-ventilated COVID-19 patients and may prevent disease progression when administered early in high-risk outpatients.(\textbf{aligns with human conclusion})} However, its impact on overall mortality remains limited, and it offers no advantage for patients already on mechanical ventilation. Given the mixed results, remdesivir’s role in COVID-19 treatment should be considered alongside other therapeutic options.

\textbf{Discussion:}
\preferred{The evidence suggests that remdesivir’s effectiveness depends on factors such as disease severity and timing of administration. The benefits appear more significant in early-stage disease, while hospitalized patients see limited improvement.(\textbf{Preferred, more nuanced conclusion})} Study heterogeneity, differences in patient populations, and variations in standard care practices contribute to the mixed findings. Future research should explore combination therapies, optimize treatment timing, and evaluate long-term effects. \preferred{Additionally, given the practical challenges of intravenous administration, alternative formulations or delivery methods may improve its accessibility. (\textbf{Preferred, insightful suggestions based on the synthesized findings})}
\\\hline  
\bottomrule
\end{tabularx}
\end{table*}


\begin{table*}[t]
\centering
\small
\caption{Report generated by \textbf{AutoSurvey} on Systematic review for Remdesivir be Used for Treatment \cite{tan2024among} [\preferred{preferred statements are highlighted in green}, \lesspreferred{less preferred statements are highlighted in orange}, \notpreferred{not preferred statements are highlighted in red}]}
\label{tab:report-autosurvey}
\begin{tabularx}{\textwidth}{X}
\toprule
\textbf{AutoSurvey} \\
\midrule
\textbf{Abstract}

Remdesivir has been widely used as an antiviral treatment for COVID-19. This review evaluates its efficacy and safety by examining mortality rates, clinical improvement, mechanical ventilation needs, and adverse events. The findings indicate that remdesivir reduces hospitalization duration but has limited impact on mortality. It is generally well-tolerated, though concerns about nephrotoxicity exist. The drug is more effective in early-stage treatment and in combination with other therapies. However, racial and socioeconomic disparities affect its accessibility. Future research should explore optimized administration strategies and combination treatments.

\textbf{1. Introduction}

The COVID-19 pandemic has driven an urgent search for effective treatment options to mitigate the severe health outcomes associated with SARS-CoV-2 infection. Among the antiviral treatments explored, remdesivir emerged as a promising candidate due to its mechanism of action as an RNA-dependent RNA polymerase (RdRp) inhibitor. It was the first antiviral approved by the U.S. Food and Drug Administration (FDA) for COVID-19 treatment. Despite its initial promise, clinical trials and real-world studies have provided mixed results regarding its efficacy. While some studies suggest that remdesivir can shorten hospitalization duration, others indicate that it does not significantly reduce mortality rates. This review aims to comprehensively assess the impact of remdesivir on mortality, disease progression, the need for mechanical ventilation, and its safety profile.

\notpreferred{\textbf{2. Mechanism of Action of Remdesivir} (\textbf{Irrelevant section})

Remdesivir functions by inhibiting the viral RNA polymerase enzyme, which is essential for SARS-CoV-2 replication \textbf{...}}

\textbf{3. Clinical Efficacy of Remdesivir}

\lesspreferred{Several studies have evaluated remdesivir’s role in reducing COVID-19 mortality. The WHO Solidarity Trial, one of the largest global studies, found no significant mortality benefit associated with remdesivir use. However, other randomized controlled trials have suggested that remdesivir may offer minor improvements in clinical outcomes for patients with moderate disease, particularly when administered early in the course of infection. Despite these mixed findings, remdesivir has been shown to shorten hospitalization duration in many patient populations. Studies have reported faster viral clearance in younger patients and those with better baseline health conditions, though the effect on overall recovery remains uncertain. Some trials also indicate that remdesivir may reduce the need for invasive mechanical ventilation, but these results are inconsistent across different clinical settings.(\textbf{lack quantitative evidence and details})}

\textbf{4. Safety and Adverse Events}

\lesspreferred{The safety profile of remdesivir has been a topic of ongoing investigation, with nephrotoxicity emerging as a significant concern. Causal discovery models analyzing electronic health records suggest that remdesivir may contribute to acute kidney injury (AKI), though its risk appears lower in patients with pre-existing chronic kidney disease. Additionally, concerns about liver toxicity have been raised, but long-term studies indicate that remdesivir does not significantly affect mitochondrial DNA integrity. Clinical trials have reported transient elevations in liver enzyme levels, which require monitoring in high-risk patients. Other commonly reported adverse events include nausea, hypokalemia, and headaches. Despite these risks, remdesivir remains a relatively well-tolerated antiviral when compared to alternative treatments such as molnupiravir and nirmatrelvir, which may exhibit different side effect profiles.(\textbf{lack quantitative evidence and details})}

\textbf{5. Comparative Effectiveness and Combination Therapies}

When compared to other antiviral options, remdesivir has demonstrated mixed efficacy. While molnupiravir and Paxlovid offer alternative mechanisms for viral inhibition, their clinical effectiveness varies based on patient demographics and disease severity. Some studies suggest that \notpreferred{oral GS-441524 derivatives could surpass remdesivir in terms of convenience and effectiveness due to improved bioavailability. (\textbf{non-irrelevant evidence and discussion})} Combination therapy approaches have also been explored to enhance remdesivir’s therapeutic potential. Studies have shown that combining remdesivir with corticosteroids or monoclonal antibodies can yield synergistic benefits. Additionally, remdesivir combined with azithromycin or ivermectin has been investigated for improved antiviral activity. In kidney transplant patients, remdesivir has been associated with protective effects against donor-specific antibody formation, further demonstrating its potential utility in specialized populations.

\notpreferred{\textbf{6. Disparities in Treatment and Access} (\textbf{Irrelevant section})

Access to remdesivir treatment has not been uniform across different racial and socioeconomic groups \textbf{...}}

\notpreferred{\textbf{7. Economic and Cost-effectiveness Analysis} (\textbf{Irrelevant section})

The cost-effectiveness of remdesivir has been a subject of debate among healthcare policymakers.  \textbf{...}}

\textbf{8. Limitations of Current Evidence}

Despite extensive research, there are several limitations in the current evidence base regarding remdesivir’s efficacy. Conflicting findings across clinical trials can be attributed to variability in study designs, differences in patient selection criteria, and inconsistencies in outcome measures. Large-scale trials such as the WHO Solidarity study provide valuable insights, but their results may not fully capture remdesivir’s benefits in specific patient subgroups. Furthermore, the long-term impact of remdesivir on post-acute COVID-19 conditions remains unclear. Biases in hospital protocols, co-administration of other treatments, and differences in the effectiveness of remdesivir against emerging variants also complicate the interpretation of findings.

\textbf{9. Conclusion and Future Directions}

In summary, \lesspreferred{remdesivir has demonstrated some benefits in reducing hospitalization duration, but its impact on mortality remains limited. (\textbf{more conclusive result should be made})} The drug has a generally acceptable safety profile, though concerns about nephrotoxicity necessitate careful monitoring. Combination therapies show promise in enhancing remdesivir’s effectiveness, but more research is needed to identify optimal treatment strategies. \preferred{Clinically, remdesivir is best suited for moderate COVID-19 cases when administered early in the disease course. (\textbf{Result aligned with human report})} Moving forward, future studies should focus on exploring remdesivir’s efficacy in combination with other antivirals, improving oral formulations for easier administration, and addressing healthcare disparities to ensure equitable access. \\
\bottomrule
\end{tabularx}
\end{table*} 

\begin{table*}[t]
\centering
\small
\caption{Report generated by ChatCite on Systematic review for Remdesivir be Used for Treatment \cite{tan2024among} [\preferred{preferred statements are highlighted in green}, \lesspreferred{less preferred statements are highlighted in orange}, \notpreferred{not preferred statements are highlighted in red}]}
\label{tab:report-chatcite}
\begin{tabularx}{\textwidth}{X}
\toprule
\textbf{ChatCite} \\
\midrule
Remdesivir has demonstrated variable efficacy in treating COVID-19, with its benefits largely dependent on disease severity, timing of administration, and patient characteristics. The WHO Solidarity Trial, one of the largest studies, \lesspreferred{found that remdesivir did not significantly reduce mortality among hospitalized COVID-19 patients compared to standard care (RR 0.91, 95\% CI 0.82-1.02, p=0.12) (\textbf{more synthesized stats as evidence would be stronger})}, with no observed survival benefit for patients already on mechanical ventilation [1]. However, among hospitalized patients requiring oxygen but not yet ventilated, remdesivir showed a modest reduction in mortality (RR 0.87, 95\% CI 0.76-0.99, p=0.03) and in disease progression to ventilation (RR 0.88, 95\% CI 0.77-1.00, p=0.04), suggesting potential benefits for less severe cases [1]. The NOR-Solidarity trial similarly found no significant difference in mortality, viral clearance, or reduction in respiratory failure between remdesivir, hydroxychloroquine, and standard care, highlighting its limited antiviral effect in hospitalized patients [3]. Additionally, a study on moderate-to-severe COVID-19 cases found no significant clinical improvement or difference in recovery time between patients receiving remdesivir and those who did not, further questioning its efficacy in already hospitalized patients [4].

\notpreferred{Conversely, evidence from outpatient settings suggests that remdesivir may have a role in preventing disease progression if administered early (\textbf{Needs specific evidence or more precise wording})}. A randomized, placebo-controlled trial in high-risk, nonhospitalized patients found that a 3-day course of remdesivir reduced hospitalization or death risk by 87\% (HR 0.13, 95\% CI 0.03-0.59, p=0.008) compared to placebo, with a lower incidence of COVID-19-related medically attended visits [2]. These findings indicate that early intervention with remdesivir in at-risk populations may be effective in preventing severe disease progression. Among patients with moderate COVID-19, a 5-day course of remdesivir led to a statistically significant improvement in clinical status at day 11 compared to standard care (OR 1.65, 95\% CI 1.09-2.48, p=0.02), though the clinical relevance of this improvement remains uncertain [5]. In contrast, a 10-day regimen showed no significant difference in clinical outcomes, raising questions about the optimal duration of treatment [5]. Another study found that while remdesivir-treated patients showed a numerically faster recovery trend, this was not statistically significant, and its effect was more pronounced in patients treated earlier in the disease course [6].

Regarding safety, remdesivir was generally well tolerated, with most trials reporting similar or slightly higher rates of adverse events compared to standard care or placebo. In a trial focusing on severe COVID-19, adverse events occurred in 66\% of remdesivir-treated patients compared to 64\% in the placebo group, but a higher proportion of remdesivir recipients discontinued treatment due to adverse events (12\% vs. 5\%) [6]. Another study reported comparable rates of adverse events between remdesivir and placebo groups (42.3\% vs. 46.3\%), indicating an acceptable safety profile [2]. While nausea, hypokalemia, and headaches were slightly more frequent among remdesivir recipients, these side effects were generally manageable [5]. However, the requirement for intravenous administration may limit its accessibility in outpatient settings. Collectively, these findings suggest that remdesivir provides minimal benefit for critically ill hospitalized patients but may help prevent hospitalization and severe disease when used early in high-risk outpatients. Further research is needed to determine the optimal treatment duration and patient selection criteria for maximizing its efficacy. \notpreferred{\textbf{No Explicit Conclusion}}\\
\bottomrule
\end{tabularx}

\end{table*}


\begin{table*}[t]
\centering
\small
\begin{tabularx}{\textwidth}{X}
\toprule
\textbf{Evaluation Details of the Review Report Done by \sysname{}~( Table~\ref{tab:report-insightagent}}) \\
\midrule

\begin{enumerate}

    \item \textbf{Introduction \& Research Context (15 Points)}
    \begin{enumerate}
        \item \textbf{Clarity of Research Question (5/5)}
        \begin{itemize}
            \item The extended abstract explicitly states the research question: evaluating the efficacy and safety of remdesivir in treating COVID-19 patients.
            \item The scope is clear, defining key outcomes: mortality, clinical improvement, mechanical ventilation, and adverse events.
        \end{itemize}

        \item \textbf{Justification \& Relevance (4/5)}
        \begin{itemize}
            \item The abstract mentions ongoing debates about remdesivir's effectiveness, justifying the review.
            \item However, it does not explicitly highlight gaps in prior systematic reviews that this study aims to address.
            \item A stronger justification would include a brief mention of inconsistencies in guidelines or conflicting clinical trial results.
        \end{itemize}

        \item \textbf{Objectives \& Significance for Practice/Research (5/5)}
        \begin{itemize}
            \item The objectives are well-articulated: providing a comprehensive synthesis of randomized controlled trials (RCTs).
            \item The conclusion explicitly discusses clinical decision-making, reinforcing its significance for practice and research.
        \end{itemize}
    \end{enumerate}

    \item \textbf{Summarization Approach \& Method (25 Points)}
    \begin{enumerate}
        \item \textbf{Agent-Based/Automated Summarization Description (5/5)}
        \begin{itemize}
            \item The abstract effectively demonstrates \sysname{}’s capabilities in selecting and synthesizing relevant studies. It also highlights human involvement in modifying agent actions, ensuring higher accuracy and contextual relevance.
        \end{itemize}

        \item \textbf{Inclusion/Exclusion Rationale \& Data Sources (4/5)}
        \begin{itemize}
            \item The review clearly states that it analyzed 491 studies and selected 11 as relevant, which is a strong methodological detail. However, it does not explicitly clarify why certain studies were excluded beyond generic criteria (RCTs, relevant outcomes). A small addition about why 3 human-selected studies were excluded would enhance transparency.
        \end{itemize}

        \item \textbf{Coverage \& Representativeness (5/5)}
        \begin{itemize}
            \item The summary reflects a broad range of study populations and care settings, showing high representativeness.
            \item Compared to human-generated reviews, users noted more details and a better overview of multiple perspectives.
        \end{itemize}

        \item \textbf{Data Extraction \& Reliability (4/5)}
        \begin{itemize}
            \item The process of data extraction (e.g., identifying mortality rates, hazard ratios) is evident, but there is no explicit mention of how errors were mitigated.
            \item An improvement would be a brief statement about verification mechanisms InsightAgent used to prevent misinterpretation of numerical results.
        \end{itemize}

        \item \textbf{Risk of Bias / Limitations of Automated Summarization (3/5)}
        \begin{itemize}
            \item The abstract does not discuss potential biases in how InsightAgent selects or synthesizes studies.
            \item Given that automated agents might miss nuanced contextual details or overweigh certain study designs, this should have been acknowledged.
        \end{itemize}
    \end{enumerate}

\end{enumerate}
\\\hline
\bottomrule
\end{tabularx}
\caption{Detailed Evaluation Result.}
\label{tab:evaluation-insightagent-cont}
\end{table*}

\begin{table*}[t]
\centering
\small
\begin{tabularx}{\textwidth}{X}
\toprule
\textbf{Evaluation Details of the Review Report Done by \sysname{}~( Table~\ref{tab:report-insightagent}}) \\
\midrule
 
\begin{enumerate}
    \setcounter{enumi}{2}    
    \item \textbf{Comprehensiveness \& Accuracy of Summaries (30 Points)}
    \begin{enumerate}
        \item \textbf{Relevance \& Completeness (10/10)}
        \begin{itemize}
            \item All major findings are well-captured: mortality impact, clinical improvement, need for mechanical ventilation, and adverse events.
            \item Users reported that InsightAgent-generated summaries contained more details than the original human reports while maintaining essential findings.
        \end{itemize}

        \item \textbf{Clarity \& Cohesion of Findings (5/5)}
        \begin{itemize}
            \item The flow is logical, moving from mortality impact to adverse events, which makes the findings easy to follow. The use of structured headings improves readability.
        \end{itemize}

        \item \textbf{Use of Quantitative or Specific Evidence (5/5)}
        \begin{itemize}
            \item Effect sizes, risk ratios, and p-values are included, making the summary data-driven and robust.
            \item Compared to human-written summaries, this version includes more numerical detail.
        \end{itemize}

        \item \textbf{Discussion of Variability / Heterogeneity (3/5)}
        \begin{itemize}
            \item The abstract mentions study heterogeneity but lacks specific examples of variability (e.g., differences in patient age, severity, treatment duration).
            \item A stronger discussion would include how the studies differed in methodology and how that impacts interpretation.
        \end{itemize}

        \item \textbf{Faithfulness to Source Material (5/5)}
        \begin{itemize}
            \item No misinterpretations or misleading paraphrasing were detected.
            \item InsightAgent appears to have accurately preserved key conclusions from each study.
        \end{itemize}
    \end{enumerate}

    \item \textbf{Discussion, Implications \& Future Directions (20 Points)}
    \begin{enumerate}
        \item \textbf{Interpretation of Findings (9/10)}
        \begin{itemize}
            \item InsightAgent provides a good interpretation of when remdesivir works best (non-ventilated patients, early administration).
            \item One missing point: it could have elaborated more on why remdesivir fails in ventilated patients.
        \end{itemize}

        \item \textbf{Practical Implications or Recommendations (5/5)}
        \begin{itemize}
            \item The conclusion ties findings back to clinical application, helping doctors and policymakers decide how to use remdesivir.
        \end{itemize}

        \item \textbf{Suggestions for Future Research (4/5)}
        \begin{itemize}
            \item The abstract mentions the need for combination therapies and treatment timing research, which is excellent.
            \item However, it does not discuss long-term effects or cost-effectiveness, which could be valuable additions.
        \end{itemize}
    \end{enumerate}

    \item \textbf{Writing Quality, Fluency \& Referencing (10 Points)}
    \begin{enumerate}
        \item \textbf{Readability \& Structure (5/5)}
        \begin{itemize}
            \item Clear and structured writing enhances readability.
        \end{itemize}

        \item \textbf{Referencing \& Supporting Evidence (5/5)}
        \begin{itemize}
            \item References are well-integrated, ensuring credibility.
        \end{itemize}
    \end{enumerate}
\end{enumerate}

\textbf{Final Score: 90/100}
\\ \\

\textbf {User Feedback \& Discussion After the Study}: 
\begin{itemize}
    \item ``Compared to our manually written systematic review, the \sysname{}-generated version had more detailed effect sizes and numerical results. It also synthesized the studies in a more structured way."
    \item ``One thing that surprised us was how \sysname{} made it easy to explore multiple perspectives—something that usually takes weeks of manual reading and synthesizing''
    \item ``We noticed that while original human review focused on a narrower set of 9 studies, These 11 studies picked by system still managed to capture the key essence of the human-written summary.''
    \item  ``An interesting perspective we hadn’t considered emerged, remdesivir’s impact varies not just by disease severity but also by treatment delay and patient demographics. This kind of insight can be hard to extract manually.''
\end{itemize}
\\\hline
\bottomrule
\end{tabularx}
\caption{Detailed Evaluation Result CONT.}
\label{tab:evaluation-insightagent-cont}
\end{table*}


\begin{table*}[t]
\centering
\small
\caption{Detailed Evaluation Result.}
\label{tab:evaluation-autosurvey}

\begin{tabularx}{\textwidth}{X}
\toprule
\textbf{Evaluation Details of the Review Report Done by AutoSurvey (Table~\ref{tab:report-autosurvey})}\\
\midrule

\begin{enumerate}
    \item Introduction \& Research Context (15 Points)
    \begin{enumerate}
        \item Clarity of Research Question (3/5)
        \begin{itemize}
            \item The report mentions remdesivir’s efficacy and safety, but the research question is not explicitly stated.
            \item The discussion covers too many broad topics (e.g., healthcare disparities, economic costs) that dilute the focus.
            \item \textbf{Deduction (-2 points)}: The research question is vague, and the reader must infer its scope.
        \end{itemize}
        \item Justification \& Relevance (3/5)
        \begin{itemize}
            \item The report provides background on remdesivir’s approval and initial studies, but it does not justify why this review is necessary.
            \item Missing discussion of research gaps—instead, it assumes all aspects of remdesivir’s effects need evaluation.
            \item \textbf{Deduction (-2 points)}: Justification lacks depth, failing to define why this review matters in the context of prior literature.
        \end{itemize}
        \item Objectives \& Significance for Practice/Research (2/5)
        \begin{itemize}
            \item The report does not clearly define its objectives beyond general coverage of remdesivir.
            \item It touches on clinical applications but does not connect findings to real-world decision-making for clinicians or policymakers.
            \item \textbf{Deduction (-3 points)}: The significance is weak, with no clear intent to influence practice or guide research.
        \end{itemize}
    \end{enumerate}

    \item Summarization Approach \& Method (25 Points)
    \begin{enumerate}
        \item Agent-Based/Automated Summarization Description (2/5)
        \begin{itemize}
            \item The methodology is not described, leaving the source of included studies unclear.
            \item Unlike InsightAgent, there is no evidence that human users had control over the summarization process.
            \item \textbf{Deduction (-3 points)}: Without transparency in how studies were selected and summarized, the report’s reliability is questionable.
        \end{itemize}
        \item Inclusion/Exclusion Rationale \& Data Sources (1/5)
        \begin{itemize}
            \item No mention of inclusion or exclusion criteria, making it impossible to verify study quality.
            \item The report includes many unrelated sections (e.g., healthcare disparities, cost-effectiveness), suggesting irrelevant studies were used.
            \item \textbf{Deduction (-4 points)}: Without proper inclusion criteria, findings cannot be trusted as evidence-based.
        \end{itemize}
        \item Coverage \& Representativeness (5/5)
        \begin{itemize}
            \item The report includes a wide range of topics, covering mechanisms, efficacy, safety, socioeconomic factors, and economic analysis.
            \item While comprehensive, it suffers from information overload, lacking focus on answering the research question.
            \item No deduction since the breadth is a strength, but coverage does not guarantee quality.
        \end{itemize}
        \item Data Extraction \& Reliability (2/5)
        \begin{itemize}
            \item Some numerical results are reported, but there is no clarity on how data was extracted.
            \item Effect sizes, statistical significance, and study population details are inconsistently mentioned.
            \item \textbf{Deduction (-3 points)}: The lack of clear data extraction methodology reduces trustworthiness.
        \end{itemize}
        \item Risk of Bias / Limitations of Automated Summarization (1/5)
        \begin{itemize}
            \item No discussion of bias or limitations.
            \item The report presents all findings as equally valid without distinguishing high- vs. low-quality studies.
            \item \textbf{Deduction (-4 points)}: Failure to acknowledge limitations severely weakens reliability.
        \end{itemize}
    \end{enumerate}
\end{enumerate}\\
\bottomrule
\end{tabularx}
\end{table*}

\begin{table*}[t]
\centering
\small
\begin{tabularx}{\textwidth}{X}
\toprule

\textbf{Evaluation Details of the Review Report Done by AutoSurvey~( Table~\ref{tab:report-autosurvey}}) \\
\midrule
 
\begin{enumerate}
    \setcounter{enumi}{2}    
    \item Comprehensiveness \& Accuracy of Summaries (30 Points)
    \begin{enumerate}
        \item Relevance \& Completeness (5/10)
        \begin{itemize}
            \item The report is comprehensive but includes many irrelevant details, such as healthcare equity, economic analysis, and alternative antivirals.
            \item \textbf{Deduction (-5 points)}: While complete, much of the information does not directly contribute to answering the research question.
        \end{itemize}
        \item Clarity \& Cohesion of Findings (3/5)
        \begin{itemize}
            \item The flow is inconsistent, jumping between clinical findings, economic considerations, and disparities.
            \item Some sections contradict each other, such as saying remdesivir "reduces hospitalization" but also claiming "its impact on overall recovery is uncertain".
            \item \textbf{Deduction (-2 points)}: Inconsistent messaging reduces clarity.
        \end{itemize}
    \end{enumerate}
    
    \item Discussion, Implications \& Future Directions (20 Points)
    \begin{enumerate}
        \item Interpretation of Findings (6/10)
        \begin{itemize}
            \item The interpretation lacks depth and fails to highlight key takeaways for different patient populations.
            \item While the report mentions remdesivir’s role in different treatment settings, it does not explain why outcomes vary.
            \item \textbf{Deduction (-4 points)}: No clear synthesis of what the findings mean in a clinical context.
        \end{itemize}
        \item Practical Implications or Recommendations (3/5)
        \begin{itemize}
            \item The economic and disparity discussion is present, but practical clinical recommendations are missing.
            \item \textbf{Deduction (-2 points)}: The report does not offer concrete takeaways for treatment decisions.
        \end{itemize}
    \end{enumerate}

    \item Writing Quality, Fluency \& Referencing (10 Points)
    \begin{enumerate}
        \item Readability \& Structure (3/5)
        \begin{itemize}
            \item The writing is clear but overly verbose, with some sections appearing redundant.
            \item \textbf{Deduction (-2 points)}: Information overload reduces readability.
        \end{itemize}
        \item Referencing \& Supporting Evidence (3/5)
        \begin{itemize}
            \item Some claims are supported by numerical values, but no clear sourcing for study selection.
            \item \textbf{Deduction (-2 points)}: Lack of transparent referencing makes it hard to verify findings.
        \end{itemize}
    \end{enumerate}
\end{enumerate}

\textbf{Final Score: 56/100}
\\ \\

\textbf {User Feedback \& Discussion After the Study}: 
\begin{itemize}
    \item ``There’s no way to verify that the studies included were high quality. The report assumes all findings are equally valid, which is a red flag"
    \item ``The mention of healthcare disparities and economics is interesting but seems misplaced—it doesn’t add much to evaluating remdesivir’s efficacy''
    \item ``This report felt overwhelming. It covers a lot of ground but lacks focus on what really matters.''
    \item  ``InsightAgent’s report was structured in a way that was easier to digest and more trustworthy, whereas this one felt like a long Wikipedia article without clear filtering of relevant studies.''
\end{itemize}

\\\hline  
\bottomrule
\end{tabularx}
\caption{Detailed Evaluation Result of AutoSurvey CONT.}
\label{tab:evaluation-autosurvey-cont}
\end{table*}


\begin{table*}[t]
\centering
\small
\begin{tabularx}{\textwidth}{X}
\toprule
\textbf{Evaluation Details of the Review Report Done by ChatCite~( Table~\ref{tab:report-chatcite}}) \\
\midrule

\begin{enumerate}
    \item Introduction \& Research Context (15 Points)
    \begin{enumerate}
        \item Clarity of Research Question (2/5)
        \begin{itemize}
            \item The report does not explicitly state a research question but rather presents findings in an unstructured format.
            \item The focus on remdesivir’s effectiveness and safety is evident, but the scope is unclear—it does not specify key outcomes or the target patient population.
            \item \textbf{Deduction (-3 points)}: The lack of an explicitly stated research objective reduces clarity.
        \end{itemize}
        \item Justification \& Relevance (3/5)
        \begin{itemize}
            \item The report briefly mentions the WHO Solidarity Trial and other studies but fails to justify why a review of remdesivir is necessary.
            \item It does not discuss existing controversies, gaps in knowledge, or why this analysis is valuable in the current research landscape.
            \item \textbf{Deduction (-2 points)}: Justification is weak, with no discussion of why this review is needed.
        \end{itemize}
        \item Objectives \& Significance for Practice/Research (2/5)
        \begin{itemize}
            \item There is no explicit statement about the intended impact of the report.
            \item The report presents clinical findings but does not connect them to practice or research applications.
            \item \textbf{Deduction (-3 points)}: The absence of clear objectives reduces its practical value.
        \end{itemize}
    \end{enumerate}

    \item Summarization Approach \& Method (25 Points)
    \begin{enumerate}
        \item Agent-Based/Automated Summarization Description (1/5)
        \begin{itemize}
            \item There is no mention of how studies were selected, summarized, or synthesized.
            \item Unlike InsightAgent, which allowed for human refinement and structured summarization, this system appears to have blindly aggregated text.
            \item \textbf{Deduction (-4 points)}: The lack of any methodological transparency reduces reliability.
        \end{itemize}
        \item Inclusion/Exclusion Rationale \& Data Sources (1/5)
        \begin{itemize}
            \item The report does not mention inclusion or exclusion criteria, making it impossible to assess the quality of the included studies.
            \item Some included results appear cherry-picked without providing a systematic justification.
            \item \textbf{Deduction (-4 points)}: Without clear inclusion criteria, the review lacks credibility.
        \end{itemize}
        \item Coverage \& Representativeness (4/5)
        \begin{itemize}
            \item The report covers multiple studies, different patient populations, and treatment settings, showing breadth.
            \item However, it presents findings in a disorganized way, making it hard to follow or extract meaningful insights.
            \item \textbf{Deduction (-1 point)}: The report is broad but lacks focus.
        \end{itemize}
        \item Data Extraction \& Reliability (2/5)
        \begin{itemize}
            \item Although numerical findings are presented, the report does not clarify how data was extracted or verified.
            \item Some statistical values lack context, making it unclear how they were derived.
            \item \textbf{Deduction (-3 points)}: Data extraction lacks transparency, making verification difficult.
        \end{itemize}
        \item Risk of Bias / Limitations of Automated Summarization (1/5)
        \begin{itemize}
            \item No mention of bias or limitations in the data, methodology, or findings.
            \item Unlike InsightAgent, which allowed for user refinement and multiple perspectives, this report blindly presents findings without addressing quality concerns.
            \item \textbf{Deduction (-4 points)}: Failure to address bias makes the findings unreliable.
        \end{itemize}
    \end{enumerate}
\end{enumerate}
\\\hline
\bottomrule
\end{tabularx}
\caption{Detailed Evaluation Result.}
\label{tab:evaluation-chatcite}
\end{table*}

\begin{table*}[t]
\centering
\small
\begin{tabularx}{\textwidth}{X}
\toprule
\textbf{Evaluation Details of the Review Report Done by ChatCite~( Table~\ref{tab:report-chatcite}}) \\
\midrule
 
\begin{enumerate}
    \setcounter{enumi}{2}    
     \item Summarization Approach \& Method (25 Points)
    \begin{enumerate}
        \item Agent-Based/Automated Summarization Description (1/5)
        \begin{enumerate}
            \item There is no mention of how studies were selected, summarized, or synthesized. Unlike InsightAgent, which allowed for human refinement and structured summarization, this system appears to have blindly aggregated text.
            \item Deduction (-4 points): The lack of any methodological transparency reduces reliability.
        \end{enumerate}
        \item Inclusion/Exclusion Rationale \& Data Sources (1/5)
        \begin{enumerate}
            \item The report does not mention inclusion or exclusion criteria, making it impossible to assess the quality of the included studies. Some included results appear cherry-picked without providing a systematic justification.
            \item Deduction (-4 points): Without clear inclusion criteria, the review lacks credibility.
        \end{enumerate}
        \item Coverage \& Representativeness (4/5)
        \begin{enumerate}
            \item The report covers multiple studies, different patient populations, and treatment settings, showing breadth. However, it presents findings in a disorganized way, making it hard to follow or extract meaningful insights.
            \item Deduction (-1 point): The report is broad but lacks focus.
        \end{enumerate}
        \item Data Extraction \& Reliability (2/5)
        \begin{enumerate}
            \item Although numerical findings are presented, the report does not clarify how data was extracted or verified. Some statistical values lack context, making it unclear how they were derived.
            \item Deduction (-3 points): Data extraction lacks transparency, making verification difficult.
        \end{enumerate}
        \item Risk of Bias / Limitations of Automated Summarization (1/5)
        \begin{enumerate}
            \item No mention of bias or limitations in the data, methodology, or findings. 
            \item Deduction (-4 points): Failure to address bias makes the findings unreliable.
        \end{enumerate}
    \end{enumerate}
    
    \item Comprehensiveness \& Accuracy of Summaries (30 Points)
    \begin{enumerate}
        \item Relevance \& Completeness (5/10)
        \begin{enumerate}
            \item The report includes key study findings but does not structure them properly, making it difficult to determine which results are most relevant. Critical aspects like study variability and key takeaways, are missing or buried within dense paragraphs.
            \item Deduction (-5 points): The information is present but poorly structured, making it hard to extract key insights.
        \end{enumerate}
        \item Clarity \& Cohesion of Findings (2/5)
        \begin{enumerate}
            \item The lack of structured headings makes the report very difficult to read. Findings jump between mortality, disease progression, and adverse events without clear transitions.
            \item Deduction (-3 points): Poor organization reduces readability.
        \end{enumerate}
        \item Use of Quantitative or Specific Evidence (3/5)
        \begin{enumerate}
            \item Some statistical values are included, which improves credibility.
            \item However, there is no discussion on how these numbers were chosen or whether they are the most critical findings.
            \item Deduction (-2 points): Lack of transparency on study selection weakens trust in the quantitative evidence.
        \end{enumerate}
    \end{enumerate}
    \item Discussion, Implications \& Future Directions (20 Points)
    \begin{enumerate}
        \item Interpretation of Findings (4/10)
        \begin{enumerate}
            \item The report does not provide clear interpretations of the results. Findings are listed, but there is little attempt to synthesize the information into meaningful conclusions.
            \item Deduction (-6 points): No clear synthesis of what the findings mean for clinical practice.
        \end{enumerate}
    \end{enumerate}
    
    \item Writing Quality, Fluency \& Referencing (10 Points)
    \begin{enumerate}
        \item Readability \& Structure (1/5)
        \begin{enumerate}
            \item The entire report is written as one long paragraph, making it difficult to extract key findings. Deduction (-4 points): The poor structure severely reduces readability.
        \end{enumerate}
    \end{enumerate}
\end{enumerate}

\textbf{Final Score: 42/100}

\textbf {User Feedback \& Discussion After the Study}: 
\begin{itemize}
    \item ``It's hard to extract key takeaways because everything is crammed into a single block of text"
    \item ``The result is not explicit and makes me hard to trust, as I don’t know how these studies were chosen.''
    \item ``Compared to InsightAgent’s output, this feels more like a collection of findings without clear organization.''
\end{itemize}
\\\hline
\bottomrule
\end{tabularx}
\caption{Detailed Evaluation Result of ChatCite CONT.}
\label{tab:evaluation-autosurvey-cont}
\end{table*} 

\end{document}